\documentclass[english]{iopart}

\usepackage{graphicx}
\usepackage{url}
\usepackage{amssymb}
\newcounter{fig}
\begin{document}

\title[Ising model and Special Geometries]
{\Large The Ising model and Special Geometries}

\author{S. Boukraa$||$, S. Hassani$^\S$, 
J-M. Maillard$^\pounds$}
\address{$||$  \ LPTHIRM and IAESB,
 Universit\'e de Blida, Algeria}
\address{\S  Centre de Recherche Nucl\'eaire d'Alger, 
2 Bd. Frantz Fanon, BP 399, 16000 Alger, Algeria}
\address{$^\pounds$ LPTMC, UMR 7600 CNRS, 
Universit\'e de Paris 6\footnote[1]{Sorbonne Universit\'es 
(previously the UPMC was in Paris Universitas).}, Tour 23,
 5\`eme \'etage, case 121, 
 4 Place Jussieu, 75252 Paris Cedex 05, France} 

\begin{abstract}
We show that the globally nilpotent $\, G$-operators corresponding to 
the factors of the linear differential operators annihilating the 
multifold integrals $\, \chi^{(n)}$ of the magnetic susceptibility 
of the Ising model ($n \, \le 6$) are  homomorphic to their adjoint. 
This property of being self-adjoint up to operator homomorphisms, 
is equivalent to the fact that their symmetric square, or their 
exterior square, have rational solutions.
The differential Galois groups are in the special orthogonal,
or symplectic, groups. This self-adjoint (up to operator equivalence) property 
means that the factor operators we already know to be Derived from 
Geometry, are special globally nilpotent operators: they 
correspond  to ``Special Geometries''. 

Beyond the small order factor operators (occurring in 
the linear differential operators associated with 
$\, \chi^{(5)}$ and $\, \chi^{(6)}$), and, in particular, those associated
with modular forms, we focus on the quite large order-twelve
 and order-23 operators.
We show that the order-twelve operator has an exterior square which annihilates
a rational solution. Then, its differential Galois group is in the symplectic group 
$\, Sp(12, \,\mathbb{C})$.
The order-23 operator is shown to factorize in an order-two operator and an order-21
operator. The symmetric square of this order-21
operator has a rational solution. Its 
differential Galois group is, thus, in the orthogonal group
 $\, SO(21, \,\mathbb{C})$.

\end{abstract}

\vskip .5cm


\noindent {\bf AMS Classification scheme numbers}: 34M55, 
47E05, 81Qxx, 32G34, 34Lxx, 34Mxx, 14Kxx 
\vskip .5cm


\section{Introduction}
\label{introduc}

In  previous papers~\cite{2008-experimental-mathematics-chi,2009-chi5,2010-chi5-exact,2010-chi6} 
some calculations have been performed on the magnetic susceptibility of the square 
Ising model, and on the $\, n$-particle contributions 
 $ \, \tilde{\chi}^{(n)}$ of the susceptibility defined by multifold integrals. 
In particular, the linear differential operator for $ \, \tilde{\chi}^{(5)}$, was 
analyzed~\cite{2009-chi5,2010-chi5-exact},
and similar calculations were carried out~\cite{2010-chi6} on $ \, \tilde{\chi}^{(6)}$. 

The small order factors occurring in the factorization  
of the linear differential operators for $ \, \tilde{\chi}^{(5)}$
and $ \, \tilde{\chi}^{(6)}$, are found to be associated
with elliptic functions, or are found to have a  
{\em modular form interpretation}~\cite{2009-global-nilpotence,2011-calabi-yau-ising}.
An order-four operator, emerging for $\, \chi^{(6)}$, is found to be associated 
to a Calabi-Yau ODE (and more precisely associated with a $\, _4F_3$ hypergeometric function with
 an {\em algebraic pull-back}).
Two linear differential operators of large orders (twelve and twenty-three) 
are too involved and have not been analyzed.

If the occurrence of linear differential operators 
{\em associated with elliptic curves} for square Ising
correlation functions, or {\em modular forms}, can be expected~\cite{2007-holonomy},
it is far from being clear what is the kind of linear differential operators that 
should emerge in quite involved {\em highly composite}
objects, like the $n$-particle components $\, \chi ^{(n)}$
of the susceptibility of the square Ising model. 
We only have the prejudice, inherited from the 
Yang-Baxter integrability of the Ising model,
that these differential operators should be "special" and could possibly be  
associated with elliptic curves\footnote[2]{Corresponding 
to the canonical parametrization of 
the Ising model~\cite{Automorphisms} in terms of elliptic functions.}. 
The result~\cite{2011-calabi-yau-ising} on the order-four operator of 
$ \, \chi^{(6)}$ which is a Calabi-Yau ODE, clearly shows that one moves
away from the elliptic curve framework, and that the Ising model 
{\em does not restrict to the theory of elliptic curves}~\cite{Kean}
(and their associated elliptic functions and {\em modular forms}).

The integrand of the multifold integrals of Ising model 
is {\em algebraic} in the variables of integration
 and in the other remaining variables.
As a consequence, these multifold integrals can be interpreted 
as {\em "Periods" of algebraic varieties} and should verify {\em globally  
nilpotent}~\cite{2009-global-nilpotence} linear differential 
equations\footnote[1]{These linear differential operators factorize 
into irreducible operators that are also necessarily
 globally nilpotent~\cite{2009-global-nilpotence}.}, i.e.  
they are~\cite{Andre5,Andre6,Andre7}
 "{\em Derived From Geometry}". In a recent paper~\cite{Christol}
we showed that the multifold integrals of Ising model 
{\em actually correspond to diagonals of rational functions}.
This remarkable property does explain, may be not the 
modularity property~\cite{2011-calabi-yau-ising} of these 
$\, n$-fold integrals, but, at least, the {\em integrality} (i.e. the 
globally bounded character) 
property of the corresponding series~\cite{Christol}.  
 
Inside this "Geometry" framework~\cite{Morrisson3},  the multifold 
integrals of Ising model seem to be even more ``selected''. 
This justifies to explore\footnote[3]{Not to be confused, 
{\em at first sight}, with the "Special Geometry" of 
{\em extended supersymmetries} that appears in string theory with moduli spaces of 
Calabi-Yau threefolds~\cite{strominger-1990}. However, the questions we address in 
this paper, and the ones in~\cite{strominger-1990} both correspond to variations 
of Hodge structures. One cannot exclude  that these two 
concepts of "Special Geometry" could be related.}  these "Special Geometries". 

Actually, in a previous paper~\cite{2013-green-sg,2013-green-sg2}, and
 with a learn-by-example approach,
we displayed a set of enumerative combinatorics examples 
corresponding to miscellaneous 
{\em lattice Green 
functions}~\cite{Guttmann-2009,Guttmann-2010,Thomas,Broadhurst-2009,Koutschan-2013,Delves,Delves2},
as well as Calabi-Yau examples, together with order-seven 
operators~\cite{DettReit-2011,Bogner-2013} associated 
with  differential Galois groups which are exceptional groups.
On the {\em irreducible} operators of these examples, 
two differential algebra properties occur simultaneously~\cite{2013-green-sg}.
On the one hand, these operators are {\em homomorphic to their adjoint}, 
and, on the other hand, their symmetric, or exterior, squares have a 
{\em rational solution}~\cite{2013-green-sg}.
These properties  are equivalent, and 
correspond to special differential Galois groups. 
The differential Galois groups are not the $\, SL(N,\, \mathbb{C})$, 
or extensions of
$\, SL(N,\, \mathbb{C})$, groups one could expect generically,
 but {\em selected} $\, SO(N, \,\mathbb{C})$,
$\, Sp(N, \,\mathbb{C})$, $\, G_2$, ... differential Galois groups~\cite{Katz}. 

An irreducible linear differential operator $\, L_q$, of order $\, q$,
 has, generically, a symmetric square ($Sym^2(L_q)$) of 
order $ \,N_s =\, q \,(q+1)/2$ and an exterior 
square ($Ext^2(L_q)$) of order $\,N_e =\, q \, (q-1)/2$.
If the Wronskian of $L_q$ is rational and $\, Sym^2(L_q)$ annihilates 
a rational solution, or is of order $\, N_s\, -1$, the group 
is in the {\em orthogonal group} $\, SO(q, \,\mathbb{C})$ 
that admits an invariant quadratic form.
If the Wronskian of $\, L_q$ is rational and $\, Ext^2(L_q)$ has a rational
solution, or is of order $\, N_e\, -1$, the group 
is in the {\em symplectic group} $\, Sp(q, \,\mathbb{C})$ 
that admits an invariant alternating form, and the order $\, q$ 
is necessarily {\em even}.

We are going to use these tools on the globally nilpotent 
operators of the $\, n$-particle 
(multifold integrals) contributions of the magnetic 
susceptibility of the Ising model,
and show that these operators are not only ``Derived from Geometry'', 
but actually correspond to ``Special Geometries''.

The paper is organized as follows. In Section \ref{recalls}, we 
recall the factorizations
of the linear differential operators corresponding 
to the linear differential equations of 
$\,{\tilde \chi}^{(5)}$ and $\,{\tilde \chi}^{(6)}$. In Section \ref{ratversusHom}, 
we show (and recall)
that all of the ``small order'' factor operators are 
{\em associated with elliptic functions}, have
a modular form interpretation, or are Calabi-Yau ODE's. These factor 
operators are, therefore,
homomorphic to their adjoints.
Sections \ref{diffopL12in} and \ref{diffopL23} are devoted 
to the ``large order'' linear differential operators 
which occur for $\, {\tilde \chi}^{(5)}$ and $\,{\tilde \chi}^{(6)}$.
Seeking homomorphisms of these ``large'' factors 
with their corresponding adjoints
is out of our current computer resources. Instead, keeping 
in mind the results of~\cite{2013-green-sg},
we look for, and produce,
the rational solutions of their exterior and symmetric squares.
Section \ref{secremarks} displays a set of (quite technical) remarks
on the subtleties of these massive calculations. Section \ref{concl} 
contains the conclusion.

\section{Recalls}
\label{recalls}

\subsection{The linear differential equation of ${\tilde \chi}^{(5)}$}
\label{subdiffeqchi5}

With series of 10000 terms (modulo a prime), we 
have obtained~\cite{2008-experimental-mathematics-chi}
the Fuchsian differential equation annihilating
 $\, \tilde{\chi}^{(5)}$, which is of order 33. 
Subsequently~\cite{2009-chi5}, it was shown that the linear combination of 
5, 3 and 1-particle contributions to the magnetic susceptibility 
\begin{eqnarray}
\label{Phi5}
\hspace{-0.95in}&& \quad  \qquad  \qquad 
 \Phi^{(5)}\,\,  = \,\,\, \, 
 \tilde{\chi}^{(5)}\,\, \,  
- \, {1 \over 2} \tilde{\chi}^{(3)}\,\, \,  +\, {1 \over 120} \tilde{\chi}^{(1)}, 
\end{eqnarray}
is annihilated by an order twenty-nine linear ODE. The corresponding linear 
differential operator $L_{29}$, factorizes as
\begin{eqnarray}
\label{L29}
\hspace{-0.95in}&& \quad \qquad  \qquad 
L_{29}\, \,  \,=\,\, \,\, \, 
L_5 \cdot L_{12}^{(\rm left)} \cdot \tilde{L}_1 \cdot L_{11}, 
\end{eqnarray}
with
\begin{eqnarray}
\label{L11}
\hspace{-0.95in}&& \quad \qquad  \qquad 
L_{11} \,\,  \,=\,\,\, \, 
 (Z_2 \cdot N_1)  \, \oplus \, V_2  \, \oplus \, 
(F_3 \cdot F_2 \cdot L_1^s).
\end{eqnarray}

The linear differential equations have been obtained in primes,
 and we have obtained 
in exact arithmetic some factors occurring in the factorization.
All the factors have been reconstructed and are known {\em in exact arithmetic},
except of $ \, L_{12}^{(\rm left)}$, $ \, L_5$ and $ \, \tilde{L}_1$, 
which are known {\em only modulo some primes}.
The linear differential operator $ \, L_5$ is irreducible.
Its analytical solution (at 0) 
has been written~\cite{2008-experimental-mathematics-chi}
 as a homogeneous polynomial of (homogeneous) degree 4
of the complete elliptic integrals $ \, K$ and $ \, E$. 

Then, considering the inhomogeneous equation
\begin{eqnarray}
\label{L24}
\hspace{-0.95in}&& \quad \qquad  \qquad 
L_{24} (\Phi^{(5)})  \,\, = \,\,\,   \, sol(L_5), 
\end{eqnarray}
where
\begin{eqnarray}
\label{factorizL24}
\hspace{-0.95in}&& \quad \qquad  \qquad 
L_{24} \,\, =\,\, \,\,   L_{12}^{(\rm left)} \cdot \tilde{L}_1 \cdot L_{11},  
\end{eqnarray}
we have reconstructed~\cite{2010-chi5-exact}, in exact arithmetic,
 $\, L_{24}$ and $\, L_5$, and have shown that $\, L_{12}^{(\rm left)}$ is 
{\em irreducible}~\cite{2010-chi5-exact}.

\subsection{The linear differential equation of $\, {\tilde \chi}^{(6)}$}
\label{diffequtildechi6}

The order-52 linear differential equation of $ \, {\tilde \chi}^{(6)}$, and the 
factorization of the corresponding linear differential operator, have been given
in~\cite{2010-chi6}. 
It was shown that the linear combination of 
6, 4 and 2-particle contributions to the magnetic susceptibility 
\begin{eqnarray}
\hspace{-0.95in}&& \quad \qquad  \qquad 
\Phi^{(6)}\,\, =\, \,\, \,  \,\tilde{\chi}^{(6)}\, \, \, 
 -  \, {\frac{2}{3}} \tilde{\chi}^{(4)}\, \, \, 
+{\frac{2}{45}}  \tilde{\chi}^{(2)},  
\end{eqnarray}
is annihilated by an order forty-six linear ODE. The corresponding
linear differential operator $\, L_{46}$, factorizes as
\begin{eqnarray}
\hspace{-0.95in}&& \quad \qquad  \qquad 
L_{46} \,\,  \,=\,\, \,\, L_6 \cdot L_{23} \cdot L_{17}, 
\end{eqnarray}
with
\begin{eqnarray}
\hspace{-0.95in}&& \quad \qquad  \qquad 
L_{17} \,\,  =\,\,\,\,   
L_4^{(4)} \oplus \left( D_x-{1 \over x} \right) \oplus  L_{3}
\oplus  \, (L_4 \cdot \tilde{L}_3 \cdot L_2), 
\end{eqnarray}
\begin{eqnarray}
\hspace{-0.95in}&& \quad \qquad  \qquad 
L_4^{(4)}\, \,= \, \,\,\,
 L_{1,3} \cdot 
 \, (L_{1,2} \oplus L_{1,1} \oplus D_x), 
 \nonumber 
\end{eqnarray}
where $\, D_x$ denotes the derivative with respect to $\, x$.

The linear differential equation has been obtained in primes,
and we have obtained, in exact arithmetic, some factor operators
occurring in the factorization.
All the factors are known {\em in exact arithmetic}, 
except of $\, L_{23}$ and \, $L_6$
which are known {\em only modulo some primes}.
While $\, L_6$ is irreducible, 
since its analytical solution (at 0) has been written
 as a polynomial expression of homogeneous degree 5 of the complete 
elliptic integrals $\, K$ and $\, E$, 
we have not reached~\cite{2010-chi6} any conclusion on whether 
the operator $\, L_{23}$ is reducible or not.
Performing the factorization based on the combination method presented
in Section~4 of~\cite{2009-chi5}, needs prohibitive computational times.

\subsection{Sum up}
\label{sumup}

We have a plenty\footnote[1]{Typically these problems can be rephrased in
terms of variation of mixed Hodge structures. To some extent, this 
explains the fact that the minimal order operators annihilating the 
$\, \tilde{\chi}^{(n)}$'s factorize in a {\em quite large number 
of factor operators}. } of linear differential 
operators occurring as factor operators in
the linear differential equations of $\, \tilde{\chi}^{(5)}$ 
and $\, \tilde{\chi}^{(6)}$.
Beyond the order-one differential operators, 
$\, L_1$, $\, N_1$ (given in~\cite{2004-chi3}), $\, L_1^s$, $\, \tilde{L}_1$ 
(given in~\cite{2009-chi5, 2010-chi5-exact}),
and the fully factorizable order-four operator 
$\, L_4^{(4)}$ (given in~\cite{2010-chi6} and its solution 
in~\cite{2005-chi4}), there are linear differential operators 
of higher orders $\, Z_2$, $\, V_2$,
$\, F_2$, $\, F_3$, $\, L_5$ and $\, L_{12}^{(\rm left)}$ for $\, \tilde{\chi}^{(5)}$,
and $\, L_2$, $ \, L_3$, $ \, \tilde{L}_3$, $\, L_4$, $\, L_6$
 and $\, L_{23}$ for $\, \tilde{\chi}^{(6)}$.

\vskip 0.1cm

The next section deals with these small order operators (up to order 6) where 
we show (and/or recall) for each one, that it is either equivalent to a symmetric power
of $\, L_E$, the differential operator corresponding 
to the complete elliptic integral $\, E$,
or has a symmetric (or exterior) square which annihilates a rational solution.
Next, it is shown that each operator $\, L_q$ is 
{\em homomorphic to its adjoint}~\cite{2013-green-sg}
\begin{eqnarray}
\label{lefthomo}
\hspace{-0.95in}&& \quad \qquad  \qquad 
 L_q \cdot \, R_n \,\,\,  = \,\,\,\,   adjoint(R_n) \cdot \,  adjoint(L_q), 
\end{eqnarray}
and we focus on the order of $R_n$, and on the coefficient in front of the higher 
derivative.
Note that, for irreducible operators, there is another equivalence relation 
\begin{eqnarray}
\hspace{-0.95in}&& \quad \qquad  \qquad 
 adjoint(L_q) \cdot S_p \,\, =\,\,\, adjoint(S_p) \cdot L_q, 
\end{eqnarray}
which sends the solutions of $\, L_q$ into the solutions of the adjoint.
In the sequel, we will consider the relation (\ref{lefthomo}).

\vskip 0.1cm

{\bf Notation:} The linear differential operators of
 $\, \tilde{\chi}^{(5)}$ and $\, \tilde{\chi}^{(6)}$
have large orders and factorize in many factors. With a few exception,
all the factor operators carry large degree polynomials whose roots 
are {\em apparent singularities}.
In the sequel we adopt the notation $\, A_n(L_q)$ for these apparent polynomials
of a linear differential operator $\, L_q$, $\, n$ denoting 
the degree of the ``apparent'' polynomial.

\section{Rational solution for $\, Sym^2$ or $ \ Ext^2$
 versus Homomorphism with the adjoint}
\label{ratversusHom}

\subsection{Rational solution for $ \, Sym^2$ or $ \, Ext^2$ }
\label{ratforSymExt}

The order-two operator $\, V_2$ (given in \cite{2009-chi5}) is equivalent
to the second order operator associated with  $ \, \tilde{\chi}^{(2)}$ (or
equivalently to $L_E$ the linear differential operator corresponding 
to the complete elliptic integral of the second kind $\, E$).
Similarly, the order-two operator $\, L_2$ (given in \cite{2010-chi6})
is equivalent to the second order operator associated with  $ \, \tilde{\chi}^{(2)}$.
This is also the case for the order-three operator $\, L_3$ (given in~\cite{2010-chi6})
which is equivalent to the symmetric square of the second order operator associated 
with  $ \, \tilde{\chi}^{(2)}$.
The equivalence occurs also for the order-five operator 
$ \, L_5$,  occurring~\cite{2009-chi5} in $\, \tilde{\chi}^{(5)}$, which 
is the {\em symmetric fourth power} of $\, L_E$, and the order-six 
operator $\, L_6$, occurring~\cite{2010-chi6} in $\tilde{\chi}^{(6)}$, which is
the {\em symmetric fifth power} of $\, L_E$. There are also the 
order-three operator $ \, Y_3$ (given in~\cite{2005-chi3-method},
 and the order-four operator~\cite{2009-global-nilpotence,2005-chi4} $\, M_2$,
which are symmetric second (resp. third) power of $\, L_E$.

For all the linear differential operators which are symmetric powers of $\, L_E$, 
their solutions are given as polynomials of homogeneous degree in the 
complete elliptic integrals. For those which are equivalent to $\, L_E$ (or
to the symmetric power of $\, L_E$), the solutions can be written as homogeneous 
polynomials  in the complete elliptic integrals
and their derivatives~\cite{2009-chi5,2010-chi6}. 

Other operators are equivalent to hypergeometric functions 
{\em up to rational pullbacks}.
The order-two operator $\, Z_2$ (given
 in~\cite{2005-connection-chi3-chi4}), occurring (also) 
in the factorization of the linear differential operator~\cite{2004-chi3}
associated with $ \, \tilde{\chi}^{(3)}$, is seen to correspond to a 
{\em modular form of weight one}~\cite{2009-global-nilpotence}.
The order-two operator $\, F_2$ (given in~\cite{2009-chi5}) corresponds to a
{\em modular form}: its solutions can be written in terms of Gauss hypergeometric
functions with pullbacks~\cite{2011-calabi-yau-ising}.

\vskip 0.1cm

Now there are linear differential operators of order $\ge \, 3$
 which are equivalent to symmetric
powers of hypergeometric functions with a pullback. 

\subsubsection{Symmetric square of $\, F_3$ \newline}
\label{symsquareF3}

The symmetric square of the order-three operator $\, F_3$ is an order-six linear 
differential operator 
which is a direct sum of an order-five and an order-one differential operators. 
The rational solution of the symmetric square of $\, F_3$, denoted 
$\, S_R(Sym^2(F_{3}))$, reads
\begin{eqnarray}
\label{RatSolsym2F3}
\hspace{-0.95in}&& \quad  \qquad  \qquad 
 S_R(Sym^2(F_{3}))\,  \,=\,\,  \,  \, 
{\frac{P_{34}(x)}{D_{26}(x) \cdot \, A_7(F_2)^2}}, 
\end{eqnarray}
with
\begin{eqnarray}
\hspace{-0.95in}&& \quad \quad  \quad 
D_{26}(x) \,\,=\,\,\,\,
 {x}^{2} \cdot \, (x-1)^{2} \, (1+2\,x)^{2} \, (4\,x-1)^{9}
 \, (1+4\,x)^{7} \, (4\,{x}^{2}+3\,x+1)^{2}, 
 \nonumber 
\end{eqnarray}
where $ \, P_{34}(x)$ is a polynomial of degree 34 given in \ref{miscellan}, 
and where the degree-seven apparent polynomial 
for the order-two operator $\, F_2$ appearing in (\ref{L11}), reads: 
\begin{eqnarray}
\hspace{-0.95in}&& \quad   \quad  \quad  \quad 
\label{appA7F2}
 A_7(F_2) \,\,= \,\,\, 
1\,\, +x\,\, -24\,{x}^{2}\, -145\,{x}^{3}\, -192\,{x}^{4}\, +96\,{x}^{5}\, +128\,{x}^{7}.
\end{eqnarray}

The irreducible linear differential operator $\,F_{3}$ 
has a symmetric square of order six 
that annihilates a rational function. The differential Galois group
of the operator $\, F_3$ is in the  orthogonal
group $ \, SO(3, \,\mathbb{C})$, or equivalently the group $\, PSL(2, \,\mathbb{C})$.
The differential operator $\,F_3$ is equivalent (up to a multiplicative function)
to the symmetric square of an order-two differential 
operator\footnote[3]{See \cite{hoeij-2007} for the reduction 
of order-three ODE to order-two ODE.}.  
The independent solutions of this order-two operator (call it $\,O_2$) are
\begin{eqnarray}
\label{O2F3}
\hspace{-0.95in}&& \quad  \quad \,  \, 
 x^{-1/3} \cdot \,\,  _2F_1\Bigl([{{1} \over {6}},\, {{1} \over {6}}],
[{{1} \over {2}}]; \, P_1 \Bigr), 
 \quad   \,  \,  \,  \,  \, 
x^{1/3} \cdot \,\,  \sqrt{P_1} \cdot \,  _2F_1\Bigl([{{2} \over {3}},\, {{2} \over {3}}],
[{{3} \over {2}}]; \, P_1 \Bigr),   
\end{eqnarray}
with:
\begin{eqnarray}
\label{pullF3}
\hspace{-0.95in}&& \quad  \quad  \, \, \, 
 P_1(x)  \, = \, \, \, \,  \,  
{\frac {1}{108}}\,{\frac { (1-4\,x^2) \, (1+ 32\,x^2)^2}{x^2}}
  \, \,  = \,  \,  \,    \, 
{\frac {1}{108}}\,{\frac { (1- 16\,x^2)^3 }{x^2}}
\,  + \, 1. 
\end{eqnarray}
The solutions of $\,F_3$ can be written in terms of the solutions 
(\ref{O2F3}) of this order-two operator $\, O_2$ 
(the order-two intertwiner $\,R_2$ 
is given in \ref{appendixSomeLn})
\begin{eqnarray}
\label{solF3}
\hspace{-0.95in}&& \quad \qquad  \quad
sol(F_3) \,\,\, =\,\,\,\,
 (1-4x)^{-9/2}\, (1+4x)^{-7/2} \cdot R_2\Bigl(sol(Sym^2(O_2))\Bigr).
\end{eqnarray}

\vskip 0.1cm

Anticipating some comments in section \ref{secremarks}, we 
give the rational solution of the {\em symmetric square of
the  adjoint} of\footnote[1]{Here $\, F_3$ is taken to be monic: its 
head coefficient of $\, D_x^3$ is normalized to $\, 1$.} $\, F_3$
\begin{eqnarray}
\label{ratsymadjF3}
\hspace{-0.95in}&& \quad  \quad \quad  \quad 
 S_R(Sym^2(adjoint(F_{3})))\, \,  \, =\,\,  \,\,\,
 {\frac{N_{33}(x) \cdot \, P_{53}(x)}{A_{37}(F_3)^2}}, 
\end{eqnarray}
with:
\begin{eqnarray}
\hspace{-0.95in}&&  \quad \quad \quad  \quad 
N_{33}(x) \, \,=\, \, \, 
x^5 \cdot \, (1-x)^2 (1-2 x)^2\,  (1+2 x)^4 \,(1-4 x)^{10}
\nonumber \\
\hspace{-0.95in}&& \qquad \qquad  \quad  \qquad \qquad  \times \, 
(1+4 x)^6 \, (1+3 x +4 x^2)^2. 
\end{eqnarray}
The degree-37 apparent polynomial $\, A_{37}(F_3)$ and the 
degree-53 polynomial $\,P_{53}(x)$ are given, respectively, in
(\ref{A37}) and (\ref{P53}).

\vskip 0.1cm

\subsubsection{Symmetric square of $ \, \tilde{L}_3$ \newline} 
\label{subsubsymtildeL3}

Similarly, the order-three differential operator $ \, \tilde{L}_3$
has a symmetric square differential operator of order-six  
which is a direct sum of an order-five and an order-one linear 
differential operators, 
the latter annihilating the rational solution
 $ \, S_R( Sym^2(\tilde{L}_{3}))$:
\begin{eqnarray}
\label{RatSolsym2tL3}
\hspace{-0.95in}&& \quad  \quad S_R(Sym^2(\tilde{L}_{3}))
\, \,\, = \,\, \,\,
{\frac {2\,-42\,x\,+225\,{x}^{2}\,-260\,{x}^{3}\,-4352\,{x}^{4}\,+49152\,{x}^{5}}
{ (1 \,- 16\,x)^{7}}}. 
\end{eqnarray}
The linear differential operator $\,\tilde{L}_3$ is equivalent
 (up to a multiplicative function)
to the symmetric square of an order-two differential operator. 
The independent solutions of this order-two operator (call it $\, O_2$) are
\begin{eqnarray}
\label{O2L3}
\hspace{-0.95in}&& \quad  \qquad \,\,\,
 _2F_1\Bigl([{{1} \over {8}},\, {{3} \over {8}}],
[{{1} \over {2}}]; \, P_1 \Bigr), 
 \qquad  
 \sqrt{P_1} \cdot \,\,   _2F_1\Bigl([{{5} \over {8}},\, {{7} \over {8}}],
[{{3} \over {2}}]; \, P_1 \Bigr),   
\end{eqnarray}
with
\begin{eqnarray}
\label{pulltL3}
\hspace{-0.95in}&& \quad   \qquad 
 P_1(x)  \,\, = \, \,\,
{\frac { (1-12\,x)^2 }{(1-16x) (1-4x)^2}}
  \,\,\,  = \,\,  \, \,
{\frac { 256 x^3 }{(1-16x) (1-4x)^2}} \,\,  \, + \, 1. 
\end{eqnarray}
The solutions of $\,\tilde{L}_3$ can be written
in terms of the solutions (\ref{O2L3}) of the order-two operator
 (the order-two intertwiner $\,R_2$ is given in \ref{appendixSomeLn}):
\begin{eqnarray}
\label{soltL3}
\hspace{-0.95in}&& \quad   \,  \, \, 
 sol(\tilde{L}_3) \, \,\,  =\,\,\,\, 
 {\frac { (1-16\,x)^{9/2} \, (1-4\,x)^{3/2}}
{{x}^{2} \cdot \, (1024\,{x}^{3}-1232\,{x}^{2}+160\,x-5) }}
\cdot \,  R_2\Bigl(sol(Sym^2(O_2))\Bigr).
\end{eqnarray}
The rational solution of the {\em symmetric square 
of the adjoint} of $\, \tilde{L}_{3}$ (taken in monic form) 
reads:
\begin{eqnarray}
\label{ratsymadjtildeL3}
\hspace{-0.95in}&& \quad  \qquad 
 S_R(Sym^2(adjoint(\tilde{L}_{3})))
\,  \, \, = \, \, \, \, \, 
{\frac{x^4 \cdot \, (1-16 x)^6 \cdot \, 
P_{10}(x)}{(1-4 x)^3 \cdot \, A_4(\tilde{L}_3)^2}}, 
\end{eqnarray}
where the degree-four apparent polynomial 
of $\, \tilde{L}_3$,  $\, A_4(\tilde{L}_3)$ reads:
\begin{eqnarray}
\label{A4tildeL3}
\hspace{-0.95in}&& \quad  \qquad 
 A_4(\tilde{L}_3)  \, \, = \, \, \, \,
4352\,{x}^{4}\,+3607\,{x}^{3}\,-1678\,{x}^{2}\,+252\,x\,\, -8. 
\end{eqnarray}
The polynomial $\, P_{10}$ is given in (\ref{P10}).

\vskip 0.1cm

\subsubsection{Exterior square of $\, L_4$ \newline}
\label{subsubL4}

The last of the ``small order'' linear differential operators is the 
order-four operator $\, L_4$ occurring
in the linear differential operator of $ \, \tilde{\chi}^{(6)}$. 
The linear differential operator $\, L_4$ has been analyzed
in~\cite{2011-calabi-yau-ising} and was shown to be 
equivalent to a {\em Calabi-Yau equation} with $\, _4F_3$
hypergeometric function with an {\em algebraic pullback}.
This algebraic pullback is simply related to the modulus 
of the elliptic functions parametrizing  the Ising model. The 
solution of this order-four operator $\, L_4$, is 
sketched in \ref{appendixSomeLnL4}.

The exterior square of $\, L_4$ is of order six. It
 is a {\em direct sum} of an order-five and 
an order-one differential operators. The order-one operator annihilates
 the rational solution, 
$ \, S_R \left( Ext^2(L_{4}) \right)$, that
 we recall~\cite{2011-calabi-yau-ising}
\begin{eqnarray}
\label{RatSolext2L4}
\hspace{-0.95in}&& \quad  \qquad \,\,\, 
 S_R(Ext^2(L_{4}))\, \,= \,\,\,\, {\frac{P_{17}(x)}
{x^9 \cdot \, (1-16 x)^{13} (1-4 x)^2 \cdot \, A_4 (\tilde{L}_3) }},
\end{eqnarray}
where $\, A_4 (\tilde{L}_3)$ is given in (\ref{A4tildeL3}). 
The degree-17  polynomial $\, P_{17}$ is given in (\ref{P17}).

The rational solution of the exterior square of the adjoint of $\, L_4$ reads:
\begin{eqnarray}
\label{extadjL4}
\hspace{-0.95in}&& \quad  
 S_R(Ext^2(adjoint(L_{4}))) \, \,=\,\, \,\,
{\frac{x^{11} \cdot \,  (1-16 x)^{14}\,  (1-4 x)^2 (1-8 x) \cdot  \, P_{17}(x)}{A_{26}(L_4)}}.
\end{eqnarray}
The degree-26 apparent polynomial $\, A_{26}(L_4)$ is given in (\ref{A26}).

\vskip .1cm

\subsection{Homomorphism with the adjoint}
\label{subHomwithadj}

All the previous linear differential operators ($V_2$, $\,L_2$, $\,L_3$, $\,L_5$, $\,L_6$) 
which are homomorphic to $\, L_E$, or homomorphic to the symmetric square of $\, L_E$, 
are naturally homomorphic with their adjoints. 
This is straight consequence of the homomorphism of $\, L_E$ with its adjoint.

The more subtle linear differential operators 
($Z_{2}$, $\,F_{2}$, $\,F_{3}$, $\,\tilde{L}_{3}$),
which have been shown~\cite{2011-calabi-yau-ising} to be
associated with {\em modular forms}, and more precisely $\, _2F_1$
hypergeometric functions up to rational (or algebraic) pull-backs,
are also homomorphic to their adjoint.
For instance, $\,Z_2$ is conjugated to its adjoint
\begin{eqnarray}
\hspace{-0.95in}&& \quad   \qquad  
Z_{2} \cdot \, W_Z(x) 
\, \,\, = \,\,  \, \,\,
W_Z(x) \cdot \, adjoint(Z_{2}), 
\qquad \quad \quad \quad \hbox{with:} \\
\hspace{-0.95in}&& \quad  \qquad  
 W_Z(x) \, \,  = \,  \,  \, \, {{(1+2\,x)\cdot (1-x) \cdot
 (96\,x^4\, +104\,x^3\,-18\,x^2\,-3\,x\,+1) } \over
 {(1\,+\,4\,x)^2 \cdot  (1\, -4\,x)^5
 \cdot (1+3\,x+4 \,x^2) \cdot x }}, 
 \nonumber 
\end{eqnarray}
where $\,W_Z(x)$ is the Wronskian of $\, Z_2$.

Similarly, $\,F_2$ is homomorphic to its adjoint
\begin{eqnarray}
\hspace{-0.95in}&& \quad  \qquad  
 F_{2} \cdot W_F(x) 
\, \,\, = \,\,  \, \, W_F(x) \cdot  adjoint(F_{2}), 
\qquad \qquad \quad \hbox{with:} \\
\hspace{-0.95in}&& \quad  \qquad  
  W_F(x) \,\,\, = \, \,\, \, {{x \cdot \, A_7(F_2) }
 \over {(1+3\,x+4\, x^2)^2 \cdot (1\, +4\, x)^3 \cdot (1\, -\,4\, x)^6 }}, 
 \nonumber 
\end{eqnarray}
where, again, $\,W_F(x)$ is the Wronskian of $\,F_2$,
and where $\, A_7(F_2)$ is the apparent polynomial of $\, F_2$ given in (\ref{appA7F2}).

\subsection{Homomorphism with the adjoint for $\, F_3$ \newline}
\label{subadjF3}

The order-three linear differential operator $\,F_3$ 
is homomorphic to its adjoint,
with a large order-two intertwiner
\begin{eqnarray}
\label{homoF3}
\hspace{-0.95in}&& \quad  \qquad  
  F_3 \cdot \, R_2 \,\, = \,\,\, adjoint(R_2) \cdot \,adjoint(F_3), 
\qquad  \qquad \quad  {\rm with}
 \nonumber \\
\hspace{-0.95in}&& \quad  \qquad  
 R_2 \,\, =\,\,\, \, a_2(x) \cdot D_x^2\,\,  + a_1(x) \cdot D_x\,\,  + a_0(x),  
\end{eqnarray}
where $\, a_2(x)$ is the {\em rational solution} of
 $\,Sym^2(F_3)$ given in (\ref{RatSolsym2F3}), and where
\begin{eqnarray}
\label{azerodex}
\hspace{-0.95in}&&    
a_1(x) \, =\,\,\,
 {\frac{P_{78}(x)}{\rho_1(x) \cdot A_7(F_2)^2 \cdot \,A_{37} (F_3)}}, 
 \quad  \, \,  
a_0(x)\, =\,\,\,
 {\frac{P_{122}(x)}{\rho_1(x) \cdot \rho_0(x) \cdot \,  A_7(F_2)^2 \cdot \, A_{37}(F_3)^2}},
 \nonumber 
\end{eqnarray}
with
\begin{eqnarray}
\hspace{-0.95in}&&   \quad 
\rho_0(x)\, \, \,=\,\, \, \, \,
x \cdot \, (x-1)  \, (16\,{x}^{2}-1)  
\, (4\,{x}^{2}-1)  \, (4\,{x}^{2}+3\,x+1),
 \nonumber \\
\hspace{-0.95in}&&   \quad
 \rho_1(x) \, \,\,=\,\, \, \, \,
{x}^{3} \cdot \, (x-1)^{3} \, (2\,x-1) 
\, (4\,x-1)^{10} \, (1+4\,x)^{8} \, (1+2\,x)^{3}
 \, (4\,{x}^{2}+3\,x+1)^{3},
  \nonumber 
\end{eqnarray}
and where $\, A_7(F_2)$ is given in (\ref{appA7F2})
and $A_{37}(F_3)$ is given in (\ref{A37}).
The various polynomials $\, P_{j}(x)$ are of degree $j$.

\subsection{Homomorphism with the adjoint for $\, \tilde{L}_3$ \newline}
\label{subadjtildeL3}

The order-three linear differential operator $ \,\tilde{L}_3$ 
is also homomorphic to its adjoint
\begin{eqnarray}
\label{homL3}
\hspace{-0.95in}&& \quad  \quad  
 \tilde{L}_3 \cdot \, R_2 \,\, \,  = \,\, \,  \,
 adjoint(R_2) \cdot \, adjoint(\tilde{L}_3), 
\qquad \qquad \quad {\rm with} \\  
\hspace{-0.95in}&&   
R_2 \,   =\,  \,  
a_2(x) \cdot  \, D_x^2 \, \, 
+ {\frac{P_{11}(x)}{(16x-1)^7 \cdot  \, \rho(x) \cdot  \,  A_4(\tilde{L}_3) }} 
\cdot  \, D_x  \, \, 
+ {\frac{P_{17}(x)}{ (16 x-1)^7 \cdot  \, \rho(x)^2  \cdot  \, A_4(\tilde{L}_3)^2}},
    \nonumber 
\end{eqnarray}
where
\begin{eqnarray}
\rho(x)  \,\, =\, \, \, x \cdot \,  (4 x-1)  \, (16 x-1), 
\end{eqnarray}
and where $ \, a_2(x)$ is the {\em rational solution} of $ \, Sym^2(\tilde{L}_3)$
 given in (\ref{RatSolsym2tL3}).

\subsection{Homomorphism with the adjoint for $\, L_4$ \newline}
\label{subadjL4}

Now, let us consider the last ``small order'' linear  differential operator 
 occurring in $ \,\tilde{\chi}^{(6)}$, namely the order-four operator $ \,L_4$.
It is also homomorphic with its adjoint with an {\em order-two} intertwiner
\begin{eqnarray}
\label{homoL4}
\hspace{-0.95in}&& \quad   \qquad  \quad 
L_4 \cdot  \, R_2 \, \,  \,   =\,  \,  \,   \, adjoint(R_2) \cdot  \,  adjoint(L_4), 
\qquad  \quad \quad \quad  {\rm with}  \\
\hspace{-0.95in}&& \quad  \qquad  \quad    
R_2  \,\, \, = \,  \, \, \,
a_2(x) \cdot \,  D_x^2 \,  \,    \,  
+ {\frac{P_{46}(x)}{ \rho_1(x) \cdot \, A_4(\tilde{L}_3) \cdot \,  A_{26}(L_4)}} 
\cdot \,  D_x 
\nonumber \\
\hspace{-0.95in}&&  \qquad  \qquad  \qquad \qquad \qquad \quad  \, \, 
+ {\frac{P_{75}(x)}{ \rho_1(x) \cdot \, \rho_0(x)
 \cdot \, A_4(\tilde{L}_3) \cdot \, A_{26}(L_4)^2}},    
\end{eqnarray}
where
\begin{eqnarray}
\hspace{-0.95in}&& \, \quad \quad  \quad \quad 
\rho_0(x)  \,\, \,  =\, \, \, \,  \, 
 x \cdot \, (4 x-1) \, (8 x-1) \, (16 x-1), \quad
\\
\hspace{-0.95in}&& \, \quad  \quad  \quad  \quad      
 \rho_1(x) \, \, \,  =\, \, \,  \,  \, 
 x^{10} \cdot  \, (4 x-1)^3 \,  (8 x-1) \,  (16 x-1)^{14},
\end{eqnarray}
where $\, A_4(\tilde{L}_3)$ is the degree-four apparent polynomial
of $\,\tilde{L}_3$ given in (\ref{A4tildeL3}), 
$\, A_{26}(L_4)$ of degree 26 is the apparent polynomial of $\, L_4$
given in (\ref{A26}), and where $ \,a_2(x)$ is 
the {\em rational solution} of $ \, Ext^2(L_4)$
 given in (\ref{RatSolext2L4}).
The polynomials $ \,P_j(x)$ are of degree $j$.

\vskip .1cm

Note that $\, R_2$, the order-two intertwiner (\ref{homoL4}), is almost self-adjoint:
it is such that 
\begin{eqnarray}
\label{Y2L}
\hspace{-0.95in}&& \quad \quad   \quad  \, \,  \, 
\,Y_2^{(L)} \, = \, \,  \,  \, R_2  \cdot \, {{1} \over {r(x)}}
 \, \, = \, \,  \,  \,  \, 
\alpha_2(x) \cdot \, D_x^2 \, \, 
+ \, \, {{ d \alpha_2(x)} \over {dx}} \cdot \, D_x \, + \, \, \alpha_0(x), 
\end{eqnarray}
is an order-two {\em self-adjoint} operator,  where $\, r(x)$ is the rational function:
\begin{eqnarray}
\label{decompL4rx}
\hspace{-0.95in}&& \quad \quad   \quad  \quad   \quad   \quad  
 r(x) \, \,\,  = \,\, \,\, 
1080 \cdot \, {{ P_{26} \cdot \,  A_4(\tilde{L}_3) } \over { 
P_{17}^2 \cdot \,  (1-8\, x)\, (1-16\,x) \cdot \, x^2}}.
\end{eqnarray}
 Since  $\,Y_2^{(L)}$ is self-adjoint, the
 Wronskian of $\,Y_2^{(L)}$ is also equal to  $\, 1/\alpha_2(x)$, 
the inverse of the head coefficient of $\,Y_2^{(L)}$. The Wronskian 
of $\,Y_2^{(L)}$ is equal to the rational function 
$\, r(x)^2 \cdot \, S_R(Ext^2(adjoint(L_{4})))$, 
where $\,S_R(Ext^2(adjoint(L_{4}))$  is (\ref{extadjL4}), the 
rational solution of the adjoint 
of $\, L_4$ (written in monic form).

The other homomorphisms of $\, L_4$ with its adjoint corresponds to
the intertwining relation
\begin{eqnarray}
\label{homoL4adj}
\hspace{-0.95in}&& \quad   \qquad  \quad \quad 
 adjoint(L_2) \cdot  \, L_4  \, \,\,   \,   = \,  \,  \,   \,  \, 
 adjoint(L_4) \cdot \, L_2, 
\end{eqnarray}
where, again, the order-two operator $\, L_2$ is almost self-adjoint:
it is such that 
\begin{eqnarray}
\label{Y2R}
\hspace{-0.95in}&& \quad \quad   \quad  \, \, \,  \,
\,Y_2^{(R)} \, \, = \, \,  \,  \, r(x)  \cdot \, L_2
\,\, = \, \,  \,  \,  \, 
\beta_2(x) \cdot \, D_x^2 \, 
+ \, \, {{ d \beta_2(x)} \over {dx}} \cdot \, D_x \, + \, \, \beta_0(x), 
\end{eqnarray}
is {\em self-adjoint},  where $\, r(x)$ is the {\em same} rational function 
(\ref{decompL4rx}). The Wronskian of $\,Y_2^{(R)}$ is 
nothing but $\, S_R(Ext^2(L_{4}))$ given by 
(\ref{RatSolext2L4}). Since  $\,Y_2^{(R)}$ is self-adjoint, the
 Wronskian of $\,Y_2^{(R)}$ is also equal to $\, 1/\beta_2(x)$, 
the inverse of the head coefficient of $\,Y_2^{(R)}$.

\vskip .1cm

{\bf Remark:} Recalling the miscellaneous 
decompositions $\, L_n \cdot L_m \, + Cst$
 (up to an overall function), obtained for the lattice Green operators 
displayed in~\cite{2013-green-sg,2013-green-sg2}, and since
 the order-four operator $\, L_4$
is homomorphic to its adjoint with order-two intertwiners 
(see (\ref{homoL4}), (\ref{homoL4adj})), it is tempting
to find such a $\, L_n \cdot L_m \, + Cst$ decomposition for $\, L_4$.
One easily deduces~\cite{2013-green-sg,2013-green-sg2} 
the following decomposition for $\, L_4$, written in monic form:
\begin{eqnarray}
\label{decompL4}
\hspace{-0.95in}&& \quad \quad   \quad  \quad   \quad   \quad   \quad  
 L_4 \, \,\, = \,\, \,\,\, 
r(x) \cdot \, (Y_2^{(L)} \cdot \, Y_2^{(R)} \, + \, 1), 
\end{eqnarray}
where $\, Y_2^{(L)}$ and $\, Y_2^{(R)}$ are the two (quite large) 
{\em self-adjoint operators} (\ref{Y2L}) and (\ref{Y2R}), and where $\,r(x)$
is the previous rational function (\ref{decompL4rx}).

\vskip .1cm

Since $\, L_4$ in (\ref{decompL4}) is monic, one has the following relation 
between the head coefficients of the two self-adjoint operators 
$\, Y_2^{(L)}$ and $\, Y_2^{(R)}$: 
\begin{eqnarray}
\label{followingrelat}
\hspace{-0.95in}&& \quad \, \,  
 r(x) \cdot \, \alpha_2(x) \cdot \, \beta_2(x) \, \, = \, \, \, 1 
 \, \quad \hbox{or:}  \quad  \quad \,   \, 
{{1} \over {\beta_2(x)}} \, \, = \, \, \, 
r(x) \cdot \, \alpha_2(x) \, \, = \, \, \, a_2(x), 
\end{eqnarray}
where $\, a_2(x)$ is the head coefficient of $\, R_2$ in (\ref{homoL4}). 

\vskip .1cm

One easily verifies that  $\, S_R(Ext^2(L_{4})) \, = \, \, 1/\beta_2(x)$
 (see (\ref{RatSolext2L4}))
is solution of the exterior square of the order-two operator $\, Y_2^{(R)}$.
The rational function $1/\alpha_2(x)$ is
solution of the exterior square of the order-two 
operator $\, Y_2^{(L)}$, we denote $\, S_R(Y_2^{(L)})$,
and is simply related to $\, S_R(Ext^2(adjoint(L_{4})))$
(see (\ref{extadjL4})): it is nothing but
 $\,r(x)^2 \cdot \,   S_R(Ext^2(adjoint(L_{4})))$. 
Actually the exterior square of the adjoint monic order-four operator 
$\, adjoint(L_4) \, = \, \, (Y_2^{(R)} \cdot \, Y_2^{(L)} \, + \, 1) \cdot \, r(x)$ 
has the same rational solution as  the exterior square of 
$\, Y_2^{(L)} \cdot \, r(x)$, namely $\, 1/r(x)^2 \cdot \, S_R(Y_2^{(L)})$.

These two results are 
a simple consequence of the fact that the rational solution of the exterior square 
of a decomposition like (\ref{decompL4}), is, in general, the rational solution of 
the right-most order-two operator in the decomposition
 (see~\cite{2013-green-sg,2013-green-sg2}). 

\subsection{Comments}
\label{subcomments}

In all these previous examples on the ``small order'' linear 
differential operators occurring
in the $\, \tilde{\chi}^{(n)}$, we have, as we showed 
for other examples in~\cite{2013-green-sg},
the simultaneous occurrence of two properties: 
the {\em homomorphism of the irreducible operator with 
its adjoint} and the {\em occurrence of a rational solution
in the symmetric square, or exterior, square} of the differential operator.
The expressions of the intertwiners given above, clearly show this link.
Each time, the linear differential operator has a symmetric square 
(or exterior square) annihilating a rational solution
(see (\ref{RatSolsym2F3}), (\ref{RatSolsym2tL3}), (\ref{RatSolext2L4})), 
it is precisely, this rational solution that appears 
in the coefficient of the higher derivative of the intertwiners 
(see (\ref{homoF3}), (\ref{homL3}), (\ref{homoL4})).
For order-two linear differential operators, the rational solution
of the exterior square is just the Wronskian.

For these small orders ($ \le 4$)  examples, one even sees that, 
for odd (resp. even) order operators, 
it is the rational solution of the symmetric (resp. exterior) square 
which builds the intertwiner. 

However, beyond these examples, for an order-four linear 
differential operator which is homomorphic to its adjoint
with an odd order intertwiner, it is the symmetric square which is involved.
\ref{genericextsym} shows the situation with {\em generic} 
linear differential operators of order three and four. 
The intertwiners of the homomorphism have rational coefficients 
only when the symmetric (or exterior)
square have rational solutions.
When there is homomorphism with the adjoint, for order-three linear 
differential operators, 
the intertwiner can be of order two, or a function, 
and it is the symmetric square which annihilates the rational 
solution (see (\ref{thea2x})).
For order-four operators, if the intertwiner is of order two, or a function,  
it is the {\em exterior square} which annihilates a rational 
solution (see (\ref{a2extorderfour})), 
while for order-one and order-three intertwiners, the rational solution 
is annihilated by the {\em symmetric square} (see (\ref{a3symorderfour})).
 
Note that we have used the results in \ref{genericextsym} 
to build the homomophisms (\ref{homoF3})
and (\ref{homoL4}) of $\, F_3$ and $\, L_4$, which are the largest 
of our small orders differential operators.

\vskip 0.2cm

We have finished with the small order factors occurring in the linear 
differential operators
of $\, {\tilde \chi}^{(5)}$ and $\, {\tilde \chi}^{(6)}$. 
The differential Galois groups of the order-three operators are 
in the  {\em orthogonal group} $\, SO(3,\,\mathbb{C})$. The differential Galois group
of the order-four operator $\, L_4$ is in the {\em symplectic group} 
$\, Sp(4, \,\mathbb{C})$. \ref{invariantform} gives examples of invariant forms 
in both cases.

\vskip 0.1cm

All the factors share many properties (global nilpotence, global boundedness, 
symplectic or orthogonal differential Galois groups) and all have been solved 
(in closed forms) in terms of elliptic integrals and modular forms.
We should note, however, that the solutions of the factors, which are not at the most right
(of the factorization), are not solution of the differential operators of the $\chi^{(n)}$.
For instance, we have the analytical solutions at the origin of both $Z_2$ and $N_1$, see (\ref{L11}),
but we have not the second analytical solution at the origin\footnote{
The second analytical solution of $Z_2 \cdot N_1$ is an integral on the solutions
of $Z_2$.}  of $Z_2 \cdot N_1$ which is a component in $\tilde{\chi}^{(3)}$ and $\tilde{\chi}^{(5)}$. 
Similarly, we have not, in closed forms, the second and third solution of
$F_3 \cdot F_2 \cdot L_1^{s}$ given in (\ref{L11}), while the solutions of $F_3$
and $F_2$ are known as hypergeometric functions with pullback.
More generally, knowing the solutions of the factors 
of the minimal order operators annihilating the $\chi^{(n)}$'s does not 
yield automatically a knowledge on the solutions (in closed forms) of these operators (except, of course,
if these factors are all in direct sum, which is not the case). To achieve a complete 
understanding of the solutions of the minimal order operators annihilating the 
$\chi^{(n)}$, some additional work remains to be done.

\vskip 0.2cm

We turn now to the ``large order'' linear 
differential operators $\, L_{12}^{(\rm left)}$ and $\, L_{23}$.
Here, and as a consequence of the large size  
of these operators, the approach for finding the intertwiners between 
$\, L_{12}^{(\rm left)}$ (and $\, L_{23}$) with their corresponding adjoints 
is hopeless with our current computational resources. In the following 
two sections, keeping in mind our results on the ``small order'' operators,
we will claim that $\, L_{12}^{(\rm left)}$ will be homomorphic to its adjoint
if we find a rational solution annihilated by the exterior square 
(or by the symmetric square). Similarly, $ \, L_{23}$ will be homomorphic 
to its adjoint, if a rational solution of the symmetric square is found.

\section{On the linear differential operator $\, L_{12}^{(\rm left)}$ in $ \, {\tilde \chi}^{(5)}$}
\label{diffopL12in}

To see whether the exterior square of $\, L_{12}^{(\rm left)}$ has a rational solution
it is simpler to start from the definition of the exterior square. 
The formal solutions (at 0) of $\, L_{12}^{(\rm left)}$ are obtained (modulo a prime) 
and are, either analytic, or logarithmic with a maximum cubic power $\, \ln(x)^3$.
The general solution (at 0) of $\, L_{12}^{(\rm left)}$ is written as 
\begin{eqnarray}
\hspace{-0.95in}&& \quad  \quad   \qquad  
F_3(x) \cdot \, \ln(x)^3 \,\, + F_2(x) \cdot \, \ln(x)^2 \,\,
 + F_1(x) \cdot  \, \ln(x) \,\, + F_0(x).
\end{eqnarray}
Some 120 starting terms are needed to generate the series $\, F_3(x)$ with 
an homogeneous recurrence and the other series $\, F_j(x)$
 with inhomogeneous recurrences.
The twelve solutions $\, S_p$ are taken as in~\cite{2010-chi5-exact}. We 
form the linear combination
\begin{eqnarray}
\label{ext2sols}
\hspace{-0.95in}&& \quad  \quad  \quad  \quad  \,  
\sum_{k,p} \, d_{k,p} \cdot \,  ( S_k \, {\frac{d S_p}{dx}} \, \, 
- S_p \, {\frac{d S_k}{dx}} ), \quad  \quad  \, \, 
k \ne \, p\, = \,\,1,\, \cdots,\,  12,  
\end{eqnarray}
which is a general solution of the order\footnote[1]{
By cancelling all the coefficients in front of each $x^n \,\ln(x)^p$ in (\ref{ext2sols}),
we find that all the $d_{k,p}$ are zero, this means that $\, Ext^2 (L_{12}^{(\rm left)})$ is
of order 66. Similarly, if we carry the same calculations for $adjoint(L_{12}^{(\rm left)})$
and find that all the $d_{k,p}$ in the corresponding combination are zero, this will mean 
that $\, Ext^2 (adjoint(L_{12}^{(\rm left)}))$ has the order 66. This calculation for 
$adjoint(L_{12}^{(\rm left)})$ has not been done.}
66 exterior square 
$\, Ext^2 (L_{12}^{(\rm left)})$.
Demanding that this combination should not 
contain log's, fixes some of the coefficients $\, d_{k,p}$.

For a rational solution of $\, Ext^2 (L_{12}^{(\rm left)})$  to exist, 
the form, which is, now, analytic at $x=\, 0$
\begin{eqnarray}
\label{formext}
\hspace{-0.95in}&& \quad \quad  \qquad  \qquad 
D(x) \cdot \, \sum_{k,p} \, d_{k,p}\cdot
  ( S_k \, {\frac{d S_p}{dx}}  \, - S_p \, {\frac{d S_k}{dx}}), 
\end{eqnarray}
should be a polynomial, and where $\, D(x)$ is the polynomial 
whose roots are the regular singularities 
of $\, L_{12}^{(\rm left)}$. Each regular singularity 
in $\, D(x)$ is taken with the power $\, n_j$, ($n_j$
being twice the maximum local exponent of that singularity in $\, L_{12}^{(\rm left)}$).

Our series $\, S_j$ are of length 600 and the coefficients depend on some
remaining $\, d_{k,p}$. By canceling the coefficients of the higher terms
in (\ref{formext}), all the coefficients down to a given term are automatically zero,
and we obtain a polynomial. The rational solution of the exterior square 
$\, Ext^2 (L_{12}^{(\rm left)})$
thus reads
\begin{eqnarray}
\label{RatSolL12right}
\hspace{-0.95in}&& \quad  \qquad \quad  \quad 
S_R(Ext^2 (L_{12}^{(\rm left)}))  \,  \,  \, = \,\, \,  \, 
{\frac{  P_{312}(x) }{A_{131}(\tilde{L}_1 \cdot L_{11}) \cdot \, D_{211}(x) }},
\end{eqnarray}
with
\begin{eqnarray}
\hspace{-0.95in}&& \quad     
D_{211}(x) \,  \,  \, = \,\, \,  \,   
x^{18} \cdot \, (2 x-1)^2 \, (x-1)^{12}\,  (x+1)^2 \,  (2 x+1)^{13}\,  (4 x+1)^{22}\,  (4 x-1)^{24}
 \nonumber \\
\hspace{-0.95in}&& \quad   \quad \quad \,\, \,  
(4 x^2-2 x-1)^2 \,  (4 x^2+3 x+1)^{14}\,  (x^2-3 x+1)^2\,   (8 x^2+4 x+1)^8
  \nonumber \\
\hspace{-0.95in}&& \quad   \quad \quad \,\, \, 
 (4 x^3-3 x^2-x+1)^6\,  (4 x^3-5 x^2+7 x-1)^8\,  (4 x^4+15 x^3+20 x^2+8 x+1)^6, 
  \nonumber
\end{eqnarray}
where $\, P_{312}(x)$ is a polynomial of degree 312,
 and where $\, A_{131}(\tilde{L}_1 \cdot L_{11})$
is the apparent polynomial of the product $\, \tilde{L}_1 \cdot L_{11}$.

The linear differential operator $\, L_{12}^{(\rm left)}$ 
is {\em irreducible}~\cite{2010-chi5-exact}.
Its exterior square annihilates a rational function. Its differential Galois group is in 
{\em symplectic group} $\, Sp(12,\,\mathbb{C})$.
Finding the rational solution of the 
{\em exterior square of the adjoint} of $\, L_{12}^{(\rm left)}$ is, for the moment,
 beyond our computer facilities. Recalling, for the order-$\, q$ 
lattice Green operators displayed in~\cite{2013-green-sg,2013-green-sg2},  
the decompositions of the type
$\, L_n \cdot L_m \, + Cst$ (up to an overall function), where 
$\, L_m$ and $\, L_n$ are  self-adjoint operators, and where
$\, m$ and $\, n$ are two integers of the same parity, it is 
tempting to imagine, for $\, L_{12}^{(\rm left)}$, a decomposition of the 
form $\, L_{2 n} \cdot L_{2 m} \, + Cst$, where the exterior square of 
 $\, L_{2 m}$ would have (\ref{RatSolL12right}) as a rational solution. 
This would imply the existence of a rational solution for the  
{\em exterior square of the adjoint} of $\, L_{12}^{(\rm left)}$,
that identifies with the rational solution of the exterior square of  $\, L_{2 n}$.

\section{On the linear differential operator $\, L_{23}$ in ${\tilde \chi}^{(6)}$}
\label{diffopL23}

To fit our scheme that the linear  differential operators, occurring in the Ising model,
correspond to "Special Geometry", the linear differential operator $\, L_{23}$ 
should have a rational solution for its symmetric square, which 
is equivalent to say that $\, L_{23}$ is homomorphic to its adjoint.

To see whether the symmetric square\footnote[1]{By cancelling all the 
coefficients in front of each $x^n \,\ln(x)^p$ and $x^a$, $a$ half integer
in (\ref{solsym2L23}), we find that all the $f_{k,p}$ are zero, which means
that $Sym^2(L_{23})$ is of order 276.}
 of $\,L_{23}$ has a rational solution,
the general solution of $\, Sym^2 (L_{23})$ is built 
from the formal solutions (mod. prime) of $\, L_{23}$ as
\begin{eqnarray}
\label{solsym2L23}
\hspace{-0.95in}&& \quad \quad  \qquad \, \, 
 \sum_{k,p} \, f_{k,p}\cdot \,  S_k \, S_p, 
 \qquad \quad k \,\, \ge\,\,  p \, \,=\,\, \, 1,\, \cdots,\, 23,  
\end{eqnarray}
which should contain neither log's, nor $\, x^a$, ($a$ half integer), 
thus fixing some of the coefficients $\, f_{k,p}$.

For a rational solution of $\, Sym^2 (L_{23})$  to exist, 
the form (analytic at $\, x = \, 0$)
\begin{eqnarray}
\hspace{-0.95in}&& \quad \quad \quad  \qquad 
D(x) \cdot \, \sum_{k,p} \, f_{k,p}\cdot \,  S_k \, S_p, 
 \qquad \quad k \,\, \ge\,\,  p \,=\,\,\,  1,\, \cdots,\, 23,  
\end{eqnarray}
should be a polynomial, where the denominator $\, D(x)$ reads
\begin{eqnarray}
\hspace{-0.95in}&& \quad \,  \quad  D(x)\, \, \,  = \,\,\,  \, \, 
x^{n_1} \cdot \, (1-16 x)^{n_2}\,  (1-4 x)^{n_3} \,  (1-9 x)^{n_4} \, (1-25 x)^{n_5} 
(1-x)^{n_6}
 \nonumber \\
\hspace{-0.95in}&& \quad \quad \qquad \,\, \,\quad \quad  \quad \quad \quad 
\times \, 
  (1-10 x+29 x^2)^{n_7} \, (1-x+16 x^2)^{n_8},
  \nonumber
\end{eqnarray}
the order of magnitude of the exponents 
$\, n_j$ being obtained from the local exponents of the singularities.

With very long series, we have found no rational solution for $\, Sym^2 (L_{23})$.
If we trust that our series are long enough, we face one of two situations. Either 
$ \, L_{23}$ is irreducible and does not follow the
 general scheme of being special geometry,
contrary to all the other operators (obtained right now) in the Ising model,
or the linear differential operator $\, L_{23}$ is reducible. 
In the latter situation (i.e. $\, L_{23}$ is {\em reducible}), 
the right factor in $\, L_{23}$ should
be of order even: let us fix this order to two. The linear differential 
operator $\, L_{23}$ is assumed
to have the factorization
\begin{eqnarray}
\label{factorizL23}
\hspace{-0.95in}&& \quad \quad  \quad \quad \quad  \quad 
L_{23} \, \,\, =\,\,\,\,   L_{21} \cdot \tilde{L}_2.
\end{eqnarray}
In this case, and assuming that the factorization (\ref{factorizL23}) is unique, 
$\, Sym^2 (L_{23})$ does not need to have a rational solution,
but its  exterior square {\em should have} an order-one right factor, since
(see Remark 7 below) 
\begin{eqnarray}
\label{extfactorizL23}
\hspace{-0.95in}&& \quad \quad  \quad  \quad  \quad \quad 
Ext^2(L_{23}) \, \, = \,\, \,  \, O_{252} \cdot \, Ext^2( \tilde{L}_2).
\end{eqnarray}

The next step is, then, to see whether it is the exterior square 
of $\,L_{23}$ which has a rational solution.
The general solution of $\, Ext^2 (L_{23})$ is written as
\begin{eqnarray}
\label{ext2solsL23}
\hspace{-0.95in}&& \quad   \qquad \, \, 
\sum_{k,p} \, d_{k,p}\cdot  ( S_k \, {\frac{d S_p}{dx}} \, 
- S_p \, {\frac{d S_k}{dx}} ), \qquad \quad 
k \ne \, p\, =\,\,1,\, \cdots,\,  23,  
\end{eqnarray}
and should not contain log's and $\, x^a$, ($a$ half integer), 
fixing some of the coefficients $\, d_{k,p}$.

For a rational solution of $\, Ext^2 (L_{23})$  to exist, 
the form (analytic at $\, x=\, 0$)
\begin{eqnarray}
\hspace{-0.95in}&& \quad  \quad \quad  \qquad \quad  
D(x) \cdot \, \sum_{k,p} \, d_{k,p}\cdot \, 
  (S_k \, {\frac{d S_p}{dx}}  \,\,  - S_p \, {\frac{d S_k}{dx}}), 
\end{eqnarray}
should be a polynomial. We, indeed, found a rational solution for $\, Ext^2 (L_{23})$
which has the form
\begin{eqnarray}
\label{RatSolext2L23}
\hspace{-0.95in}&& \quad   \quad  \qquad \quad \quad  \qquad 
{\frac{A_{79}(L_{17}) \cdot \,   P_{93}(x)}{
x^{4} \cdot \,  (1-16 x)^{6}}}, 
\end{eqnarray}
where $\, P_{93}$ is a degree 93 polynomial, and where $\, A_{79}(L_{17})$
is the degree-79 apparent polynomial of $\, L_{17}$.

The existence of a rational solution means that an invariant alternating form
is preserved, and this implies that the order is even. Since, this is not the case,
the differential operator $\, L_{23}$ has a right factor of even order.
To establish the factorization (\ref{factorizL23}), we should show that the rational 
solution (\ref{RatSolext2L23}) is for the exterior square of an {\em order-two} linear 
differential operator.

\subsection{The factorization $\, L_{23} \,=\, L_{21} \cdot \tilde{L}_2$} 
\label{subfactoL23}

The rational solution of $\, Ext^2( \tilde{L}_2)$ does not 
determine the linear differential operator
$\, \tilde{L}_2$ whatever its order is. 
We show now, how we have obtained the linear differential operator $\, \tilde{L}_2$.

\vskip .1cm 

Once the combinations of the $\, d_{k,p}$ (in (\ref{ext2solsL23}))
 in front of the log's and $\, x^a$, ($a$ half integer) 
have been canceled, the rational solution is looked for. This fixes all the $\, d_{k,p}$.
We collect (\ref{ext2solsL23}) over $\, d\,S_j/dx$, i.e. we consider 
the linear combination of series in front 
of $\, d\,S_j/dx$. Here $\, S_j$ is the solution with the maximum power of $\, \ln(x)$,
 which is $\, \ln(x)^4$ for $\, L_{23}$.
We get no combination. There is also no combination when we collect over 
$\, d\,S_j/dx$, with $\, S_j$
the solution with $\, \ln(x)^3$ and $\, \ln(x)^2$ terms.
When $\, S_j$ is the solution with a $\, \ln(x)$ term, we obtain six identical 
combinations, i.e. series. 
If there is a right factor in $\, L_{23}$, this right factor will be of order $\ge\,  2$. 
We then search the linear ODE, that annihilates the combination found, 
and find that its order is, in fact, {\em two},
this is $\, \tilde{L}_2$.

\vskip .1cm 

We have then shown that $\, L_{23}$ actually has the factorization (\ref{factorizL23}).
Acting by $\, \tilde{L}_2$ on the solution of $\, L_{23}$ gives a series which is used
to obtain $\, L_{21}$.

The linear ODE corresponding to $\, \tilde{L}_2$ has the (analytic at $\, x=\, 0$) solution
\begin{eqnarray}
\label{soloftL2}
\hspace{-0.95in}&& \quad  \quad  \quad  \quad   \quad \quad   
{\frac{(1-16x) \cdot \, P_{90}(x) \cdot \, K(x) \, \,  \, 
 + P_{91}(x) \cdot \, E(x)}{x^{13} \cdot \, (1-16x)^{15} \, (1-4x)^{2} \, (1-8x) \cdot \, A_{79}(L_{17})}}, 
\end{eqnarray}
where $\, P_{90}$ and $\, P_{91}$ are polynomials of degree 90 and 91,
 and where $\, K(x)$ and $\, E(x)$ are the complete elliptic integrals
of the first and second kinds, 
$\, K(x)\, = \, \, _2F_1([1/2, 1/2], [1], 16x)$,
$\, E(x)\, = \,\,  _2F_1([1/2, -1/2], [1], 16x)$.
The linear differential operator $\, \tilde{L}_2$ is 
equivalent to $\, L_E$ the differential operator
of the elliptic integral $\, E(x)$. The solution (\ref{soloftL2}) 
is for $\, \tilde{L}_2$ appearing in 
the factorization $\, \tilde{L}_2 \cdot\,   L_{17}$, where 
all the linear differential operators are {\em monic}, 
and of minimal orders.

The rational solution of $\, Ext^2( \tilde{L}_2)$ reads
\begin{eqnarray}
\label{RatSolext2tL2}
\hspace{-0.95in}&& \quad   \qquad  \qquad  
{\frac{P_{93}(x)}
{x^{24} \cdot \, (1-16x)^{26} \, (1-4x)^{4} \, (1-8x)^2 \cdot \, A_{79}(L_{17})}}, 
\end{eqnarray}
which is, as it should, the rational solution (\ref{RatSolext2L23}), divided by the square of 
the polynomial in front of the higher derivative of $ \,L_{17}$.
Having obtained the linear differential operator  $ \,\tilde{L}_2$, we can see that the
roots of the polynomial $ \,P_{93}(x)$ are {\em apparent singularities} of $ \,\tilde{L}_2$,
$\, A_{93}(\tilde{L}_2) \, = \, \, P_{93}(x)$.

We turn now, to the linear differential operator $ \,L_{21}$,
 where the same calculations (as for $ \,L_{23}$) 
are performed on its {\em symmetric square}. We find that
 $ \, Sym^2 (L_{21})$ has a rational solution which reads
\begin{eqnarray}
\label{RatSolL21}
\hspace{-0.95in}&& \quad \qquad 
  S_R(Sym^2(L_{21})) \, \,  \,\,  =\,\, \, \, \, {\frac{P_{714}(x)}{D_{529}(x)}},
  \\
\hspace{-0.95in}&& \quad \qquad 
  D_{529}(x) \, \,=\,\,  \, \,
 x^{13} \cdot \,  (1-16x)^{56}\,   (1-4x)^{63} \,   (1-9x)^{47} \,  (1-25x)^{63} \,  (1-x)^{47},
 \nonumber \\
\hspace{-0.95in}&& \quad  \quad   \qquad  \qquad \quad \qquad \times \, 
(1-10x+29x^2)^{57} \,  (1-x+16x^2)^{63}, 
\nonumber
\end{eqnarray}
where $\, P_{714}$ is a polynomial of degree 714.

With the assumption that $\, L_{21}$ is irreducible (see Remark 8, below), 
the differential Galois group of $\, L_{21}$ is seen to be included
 in the orthogonal group $\, SO(21,\,\mathbb{C})$.

\vskip 0.1cm

\section{Remarks} 
\label{secremarks}

In this section we give some technical remarks on the computations 
displayed in Section \ref{diffopL12in} and Section \ref{diffopL23}. The way, the 
operator $ \, \tilde{L}_2$ has been 
obtained, is applied to show that $ \, L_{21}$ is irreducible.
By simple arguments based on the number of logarithmic solutions
of maximum degree occurring in $\, L_{21}$, we exclude the possibility
that $ \, L_{21}$ can be a symmetric power of an operator of smaller order.

\vskip 0.1cm

{\bf Remark 1:} 
From the factorization (\ref{factorizL24}) we obtained $\, L_{12}^{(\rm left)}$ by 
right division of $ \, L_{24}$ by the order-twelve operator
 $ \, \tilde{L}_1 \cdot \, L_{11}$ in its 
non monic form, because this is more tractable. 
This means that the linear differential operator $\, L_{12}^{(\rm left)}$ we are using is in fact
$\, L_{12}^{(\rm left)} \cdot \, P_{12}^{(\rm right)}$, where $ \, P_{12}^{(\rm right)}$ is the
polynomial in front of the derivative $ \, D_x^{12}$ of $ \, \tilde{L}_1 \cdot \, L_{11}$.
To obtain and give the rational solution (\ref{RatSolL12right}), we have corrected
by dividing by the square of $\, P_{12}^{(\rm right)}$. 
The rational solution (\ref{RatSolL12right}) is the solution we would have obtained
if we have used the "exterior power" then "ratsols" commands of DEtools in Maple,
on the differential operator $\, L_{12}^{(\rm left)}$ in the factorization 
$\, L_{12}^{(\rm left)} \cdot \, \tilde{L}_1 \cdot\,  L_{11}\, $
 with $\, \tilde{L}_1 \cdot\,  L_{11} \, $ 
in monic form.

\vskip 0.1cm

{\bf Remark 2:} Another remark is that, we can safely use (and we did) a non-minimal 
order (in this case order 37) ODE for $ \, L_{12}^{(\rm left)}$. 
The minimal order $\, L_{12}^{(\rm left)}$ has at the higher derivative,
besides the regular singularities, the polynomials 
$\, A_{131}(\tilde{L}_1 \cdot L_{11})^{11} \cdot \,  A_{828}(L_{12}^{(\rm left)})$, 
where $ \, A_{828}(L_{12}^{(\rm left)})$ is the degree 828 
apparent polynomial of $\,L_{12}^{(\rm left)}$ . 
The whole coefficient is of degree 2317, to be compared with the degree of the coefficient 
in front of the higher derivative $\, D_x^{37}$ of the non-minimal ODE which is 160.
At the formal solutions generation step, there are only twelve solutions that appear. 
This is because, at the point $\, x =\, 0$, all the extra and spurious solutions correspond 
to critical exponents {\em that are not, in general, rational numbers}.
In the modulo prime calculations, these exponents appear 
as roots of polynomials of degree 2 and higher. When the spurious exponent appears integer,
we should redo the calculations with another prime, or change to another non minimal
order equation.

\vskip 0.1cm

{\bf Remark 3:} One remarks that the rational solution given in (\ref{RatSolL21})
has not been corrected (as we did 
in (\ref{RatSolL12right}), (\ref{RatSolext2tL2}), see Remark 1)
by dividing by the square of the coefficient of
 higher derivative of $\, \tilde{L}_2 \cdot \, L_{17}$, which
reads
\begin{eqnarray}
\hspace{-0.95in}&& \quad  \quad  \quad \quad   
x^{12} \cdot \, (1-8 x)\,  (1-4 x)^2 \, (1-16 x)^{11}\, 
 \cdot \, A_{79}(L_{17})^2 \cdot \, A_{93}(\tilde{L}_2).
\end{eqnarray}
This type of correction is done when, in a given factorization,
 we deal with non monic factors 
which is more tractable for large operators. 
In the case of the rational solution (\ref{RatSolL21}), 
this corresponds to $ \, L_{21}$ annihilating
the series obtained by acting with a {\em non minimal}
 order-23 $ \, \tilde{L}_2$ on the solution
of $\, L_{23}$. To correct (\ref{RatSolL21}),
 as we did for (\ref{RatSolL12right}) and (\ref{RatSolext2tL2}),
we should have obtained $L_{21}$ by using 
the $ \, \tilde{L}_2$ in {\em minimal order}.
In this case, the length of the series to encode $ \, L_{21}$ is very high.
However, the occurrence of a rational solution 
to the symmetric square of $ \, L_{21}$ 
can be seen whether we use a minimal order ODE,
 or a non minimal order ODE\footnote[1]{See \ref{appendixF3F2},
 and~\cite{2008-experimental-mathematics-chi},
for a deeper understanding 
of the non minimal order representation of an operator.} for $ \, \tilde{L}_2$. 
\ref{appendixF3F2} shows the results on $ \, Sym^2(F_3)$ 
in the factorization $ \, F_3 \cdot F_2$,
which occurs in the linear differential operator of $ \, \tilde{\chi}^{(5)}$ in both ways, 
i.e. the rational solution of $ \, Sym^2(F_3)$, 
where $\, F_3$ is obtained from the factorization 
$\, F_3 \cdot \,  F_2$ with $\, F_2$ of minimal order and with $\, F_2$ of non minimal order.

\vskip 0.1cm

{\bf Remark 4:} The rational solution of the
 symmetric (or exterior) square of the operator
$\, L_q$ may carry some or all the regular
 singularities of $\, L_q$. In the expressions
of the rational solutions given
 in (\ref{RatSolsym2F3}), (\ref{RatSolext2L4}), (\ref{RatSolL12right})
and (\ref{RatSolext2tL2}), one remarks that some {\em apparent polynomials}
 occur.
These polynomials are apparent for the operator
 which is {\em at the right} of $ \, L_q$. For the operator
$\, L_q$, these polynomials are poles (see Appendix B.1 in~\cite{2009-chi5}).
For the rational solution given in (\ref{RatSolsym2tL3}) there 
is no such apparent polynomial
because the factor $\, L_2$, at the right of $\, \tilde{L}_3$, has 
no apparent singularities.
Note that the apparent polynomial of the operator,
 itself, appears in the denominator of the rational solution of the symmetric 
(or exterior) square when we deal with the adjoint of the operator.
If we call $\, \rho_j$ the local exponents at a regular (or apparent) 
singularity of $\, L_q$,
at the same point the local exponents of
 $ \, adjoint(L_q)$ are $ \, -\rho_j \, + q \, -1$ (and
$\, -\rho_j\,  - q\,  +1$ for the singularity at infinity).
An apparent singularity has the local exponents 
$\, 0, \, 1, \, \cdots,\,  q-2,\,  q$, for the adjoint
one gets automatically a pole. 

\vskip 0.1cm

{\bf Remark 5:} For all our operators $\, L_q$ we have obtained a rational solution 
for the symmetric (or exterior) square which have the maximum order 
$ \, N_s= \, q(q+1)/2$ (or $\, N_e=\, q(q-1)/2$).
For irreducible operator $\, L_q$ with differential Galois group
 in $ \, SO(q,\, \mathbb{C})$ (resp. $ \, Sp(q,\, \mathbb{C})$),  
it may happen that $\, Sym^2(L_q)$ (resp. $\, Ext^2(L_q)$) does not have 
the generic order, but has the order $ \, N_s \, -1$ (resp. $ \, N_e \, -1$). 
This means that there is no rational solution but there is, instead, 
a relation between the solutions of $\, L_q$ (or the solutions of $ \, L_q$ and 
their first derivative). Whether this drop in the order of $ \, Sym^2(L_q)$ 
(resp. $\, Ext^2(L_q)$) will also be seen for $ \, Sym^2(adjoint(L_q))$ 
(resp. $ \, Ext^2(adjoint(L_q))$), depends on the intertwiner 
occurring in the homomorphism (\ref{lefthomo}) (details will be given elsewhere).

Recall that we have not obtained a definitive conclusion on whether the exterior square 
(resp. symmetric square) of the adjoint of $\, L_{12}^{(\rm left)}$ (resp. $\, L_{21}$) 
has a drop in its order or annihilates a rational solution.

\vskip 0.1cm

{\bf Remark 6:} We have succeeded to factorize the linear differential 
operator $ \, L_{23}$ via the rational solution of $ \, Ext^2(L_{23})$. 
This, then, completes the factorization method
 we forwarded in section 4 of~\cite{2009-chi5}. 
Recall that, in this method, we produce the general (analytic at 0)
solution of $L_{23}$ which begins as
\begin{eqnarray}
\label{formalL23}
\hspace{-0.95in}&& \quad  \quad \quad  \qquad  
a_0 \,  + a_1 \, x  \, + a_2 \, x^2  \, + a_3 \, x^3  \,
 + a_4 \, x^4  \, + a_5 \, x^5  \, \,  +  \, \cdots
\end{eqnarray}
with the higher coefficients depending on the $\, a_j$, ($j= \,\,  0,\, \cdots,\,  5$).
We let the coefficients vary in the range $\, [1,\,  p_r]$, $\, p_r$ being the prime,
until a differential equation of order less than 23 is found.
If this happens, there is a right factor to $\, L_{23}$. This computation should 
have required the maximum time of $ \, 2 \, p_r^5 \,t_0$ as necessary
to produce $ \, \tilde{L}_2$, if  $\, t_0$ denotes the time needed to obtain $ \, L_{23}$.
The $ \, d_{k,p}$ coefficients, we mentioned in paragraph before (\ref{soloftL2}), are
precisely the actual values of the $a_j$ in (\ref{formalL23}) for which
the series (\ref{formalL23}) will be solution of an order-two ODE.

\vskip 0.1cm

{\bf Remark 7:} Note that (\ref{factorizL23})
 and (\ref{extfactorizL23}) are an obvious property
of the exterior power, which states that if
 $ \, L_q =  \, L_{q-n} \cdot L_n$, then the exterior power
$Ext^n \left( L_q \right)$ will have the order-one right factor
\begin{eqnarray}
\hspace{-0.95in}&& \quad  \quad  \qquad \quad  
Ext^n(L_n) \,\,\, = \,\,\,\, \,  
D_x \,\,  \, - {\frac{d}{dx}} \, \ln(W(x))
\end{eqnarray}
where $\, W(x)$ is the Wronskian of $ \, L_n$.
For our purposes, it happens that
 the suspected right factor (i.e. $ \, \tilde{L}_2$) is of 
{\em order two}, and we are dealing with the {\em second} exterior power.
If our suspected right factor in $\, L_{23}$ were not of order two, 
we would still use
\begin{eqnarray}
\hspace{-0.95in}&& \quad \qquad \quad  \quad  
 Ext^2(L_{23-n} \cdot L_n) \,\,\,  =\,\,  \,\,
 O_{253-n(n-1)/2} \cdot Ext^2 \left( L_n \right)
\end{eqnarray}
which is a general identity and we would expect 
the rational solution of $ \, Ext^2(L_{23})$ 
to come from $ \, Ext^2(L_n)$ and use the recipe (paragraph before (\ref{soloftL2})) 
to obtain $ \, L_n$.

\vskip 0.1cm

{\bf Remark 8:} 
The general (analytic at 0) solution of $L_{21}$ begins as
\begin{eqnarray}
\label{formalL21}
\hspace{-0.95in}&& \quad \quad  \quad  \qquad 
 b_1 \, x \, \, + b_2 \, x^2\, \,  + b_3 \, x^3 \, \, 
+ b_4 \, x^4 \,\,  + b_5 \, x^5\,\,\,   +\,  \cdots
\end{eqnarray}
with the higher coefficients depending on the $\, b_j$, ($j= \,\,  1,\cdots, 5$). 
We have then 
four coefficients to vary, which is very time consuming.
The way we have factorized $\, L_{23}$ can be repeated for $\, L_{21}$.
Here also, once the coefficients in $\, \sum f_{k,p} \,S_k\,S_p $ 
have been fixed to encode the 
rational solution (\ref{RatSolL21}), we collect over the series $\, S_j$,
 which is in $\, \ln(x)^4$.
We obtain two series ($L_{21}$ has two series in $\, \ln(x)^4$, see Remark 9).
If there is a right factor to $\, L_{21}$, it should be of 
order $\ge\,  5$. If it exists, its solution is a combination of both series.
This way, we have reduced the ODE search from varying four 
coefficients to one coefficient.
The computation time is still high, but the calculation can be done 
in parallel on many subintervals of $\, [1,\,  p_r]$.
We find that for {\em any} combination the result is an ODE of order 21.
This means, that the order-21 differential operator $\, L_{21}$ is
{\em irreducible}.

\vskip 0.1cm

{\bf Remark 9:} 
A last remark on the irreducibility of the large order linear differential operators
$\, L_{12}^{\rm (left)}$ and $\, L_{21}$ is worthy to be mentioned.
The operator $\, L_{12}^{\rm (left)}$ has been proved to be irreducible
 in~\cite{2010-chi5-exact}. We also showed, in~\cite{2010-chi5-exact}, that {\em it is not} 
a symmetric power, or a symmetric product, of {\em smaller order operators}
(see section 3.1 of~\cite{2010-chi5-exact}). 
We address the same issue on $\, L_{21}$, which even if it is
irreducible, it can well be built from factors of lower order, as
a symmetric power. The $n^{th}$ symmetric power of the generic order$-q$ 
operator $\, L_q$ is $\, (q+n-1)!/(n! (q-1)!)$. 
For operators $\, L_{21}$ and $ \, Sym^n(L_q)$ to be equivalent, where the doublet  
$(q, n)$ are in the only possibilities 
$\, (2,\,  20)$, $\, (3,\,  5)$ and $\, (6,\,  2)$,
their singular behavior at any singular point should match.
The linear differential operator $\, L_{21}$ has the same structure of solutions as
the operator \, $L_{23}$ (see section 4.3 of~\cite{2010-chi6}), except of 
one analytical (at the origin) and one logarithmic solutions which are solutions
of the right factor $\, \tilde{L}_2$. The local structure of the formal solutions 
(around the origin) of $\, L_{21}$ can be grouped as the following.
There are two sets of five solutions, behaving
 as $\ln(x)^k$, $\, k=\, 0, \, \cdots,\,  4$, for each set.
There are three sets of three solutions behaving 
as $\, \ln(x)^k$, $\, k=\, 0,\,  \cdots,\,  2$, for each set.
Finally, two non-logarithmic solutions behaving as $\, x^{-11/2}$ and $\, x^{-13/2}$.
For the doublets $(q,n)$, there is no possibility to obtain
 {\em two solutions} with $\, \ln(x)^4$.
The linear differential operator $\, L_{21}$ {\em is not a 
symmetric power of an operator
of smaller order}.

\section{Conclusion}
\label{concl}

This work gives a final completion of previous studies on the {\em factors}
 of the linear differential
operators associated with the $n$-particle contributions to the 
magnetic susceptibility of the Ising model (up to $n=6$).

We have shown that the globally nilpotent $\, G$-operators
corresponding to the small order ($\le \, 6$) factor operators of
the linear differential operators annihilating 
the multifold integrals $\, \chi^{(n)}$,
associated with the $\, n$-particle contributions 
 of the magnetic susceptibility
of the Ising model ($n \, \le \, 6$), are  {\em homomorphic to their
adjoint}.
This ``duality'' property of being self-adjoint up to operator
homomorphisms, is equivalent to the fact that their symmetric 
(or exterior) square have rational solutions~\cite{2013-green-sg}.
These operators are in selected differential 
Galois groups like $\, SO(q, \,\mathbb{C})$
and $\, Sp(q, \,\mathbb{C})$.
This self-adjoint (up to operator equivalence) property 
means that the factor operators, we already know
to be Derived from Geometry, are ``special''
globally nilpotent operators: they correspond
 to ``Special Geometries''. 

Two large order operators occur in the factorization of the linear differential
operators associated to $\,  \chi^{(5)}$ and $\, \chi^{(6)}$.
The order-twelve operator $\,L_{12}^{\rm left}$ has an {\em exterior square} that
annihilates a rational solution, and the order-21 operator $L_{21}$ has 
a {\em symmetric square} which annihilates a rational solution.
The different differential Galois groups
 are respectively in the symplectic group $\,  Sp(12, \,\mathbb{C})$ and
the orthogonal group $\,  SO(21, \,\mathbb{C})$.

The two properties (homomorphism with the adjoint and occurrence of a 
rational solution for the symmetric, or exterior, square), 
should be verified for these large order operators $\, L_{12}^{\rm left}$ and $\, L_{21}$.
Unfortunately, seeking for an homomorphism between these operators, and the corresponding
adjoint, is well beyond the possibility of the present computer facilities.
One may just imagine that this can be doable with dedicated programs,
computing the homomorphism\footnote[2]{For such ``massive'' formal calculations 
switching to the linear differential systems associated with these operators, 
is probably a way to calculate these homomorphisms. One first obtains the
rational solutions of the symmetric or exterior square of these differential 
systems. The intertwiners are, then, deduced from these rational solutions.} 
modulo primes, with the knowledge that the coefficient of the
higher derivative of the intertwiner is\footnote[1]{See, 
for instance, relations (\ref{thea2x}) and (\ref{a2extorderfour}).} 
the rational solution we obtained (see (\ref{RatSolL12right}) and (\ref{RatSolL21})). 
In view of the results of \ref{genericextsym}, of many examples, 
and of the examples on the five-dimensional (and six-dimensional) face-centered cubic
lattice Green function~\cite{Broadhurst-2009, Koutschan-2013}
 (namely $\, G_6^{5Dfcc}$ and $\,G_8^{6Dfcc}$ in~\cite{2013-green-sg},
see the intertwiners in equations (40) and (60) in~\cite{2013-green-sg}), 
one may concentrate on intertwiners with {\em even} orders. 
It is thus challenging to obtain the intertwiners occurring in the homomorphisms of
$\,L_{12}^{\rm left}$ and $\, L_{21}$ with their corresponding adjoint, and see whether a
"decomposition" (see equation (68) in~\cite{2013-green-sg}) 
in terms of the intertwiners occurs, the "decomposition" being probably more complex.

Without waiting this ``consolidation'', we may conclude on the intriguing selected 
character of the globally nilpotent operators annihilating the multifold integrals 
of the Ising model which are all {\em diagonals of rational functions}~\cite{Christol}, 
and we may conjecture that {\em all} the factors occurring in the differential
operators for $\chi^{(n)}$ (any $n$) {\em correspond to selected differential Galois groups}.

\ack

We thank  Y. Andr\'e, D. Bertrand, for fruitful discussions on differential 
Galois groups and (self-adjoint) dualities in geometry. 
We thank A. Bostan, G. Christol, P. Lairez and J-A. Weil for exchange of mails 
and discussions. This work has been performed without
 any support of the ANR, the ERC or the MAE.

\appendix

\section{Miscellaneous polynomials occurring in the main formulae of the paper}
\label{miscellan}

\vskip .1cm 

\subsection{The polynomials $\, A_{37}(F_3)$, $\, P_{34}$ 
and $\, P_{53}$  for $\, F_3$ \newline}
\label{miscellF3}

The apparent polynomial $\, A_{37}(F_3)$ of the (monic) order-three operator $\, F_3$,
occurs in the expression of the rational solution (\ref{ratsymadjF3})  of the 
symmetric square of the adjoint of (the monic order-three operator) 
$\, F_3$, as well as in (\ref{azerodex}).
This degree-27 polynomial reads:
\begin{eqnarray}
\label{A37}
\hspace{-0.95in}&&  \, \quad 
A_{37}(F_3)  \, \, \, = \, \, \,
5629499534213120\,{x}^{37} \, +5348024557502464\,{x}^{36} \,
 -62874472922742784\,{x}^{35} 
\nonumber \\ 
\hspace{-0.95in}&&  \, \quad  \quad 
\, +339080589913096192\,{x}^{34} \, 
+132348214635397120\,{x}^{33} \,+354600746294968320\,{x}^{32} \,
\nonumber \\ 
\hspace{-0.95in}&&  \, \quad  \quad 
+1383732497338073088\,{x}^{31} \,-269118080922157056\,{x}^{30} \,
-1021414905992970240\,{x}^{29} \,
\nonumber \\ 
\hspace{-0.95in}&&  \, \quad  \quad 
+401943021895024640\,{x}^{28} \,+378516473892569088\,{x}^{27} \,
-379126125978189824\,{x}^{26} \,
\nonumber \\ 
\hspace{-0.95in}&&  \, \quad  \quad 
-181955521970962432\,{x}^{25} \,
+182991453503356928\,{x}^{24} \,+119809766351437824\,{x}^{23}
\nonumber \\ 
\hspace{-0.95in}&&  \, \quad  \quad 
-34528714733649920\,{x}^{22} \,
-46719523456286720\,{x}^{21} \,-1865897472688128\,{x}^{20} \,
\nonumber \\ 
\hspace{-0.95in}&&  \, \quad  \quad 
+9861412040736768\,{x}^{19}
 \,+1690374175916032\,{x}^{18} \,
-1285664678690816\,{x}^{17} \,
\nonumber \\ 
\hspace{-0.95in}&&  \, \quad  \quad 
-304716171767808\,{x}^{16} \,
+112170181177344\,{x}^{15} \,+30517814178816\,{x}^{14} \,
\nonumber \\ 
\hspace{-0.95in}&&  \, \quad  \quad -7815766123264\,{x}^{13} \,
-2274047571904\,{x}^{12} \,+456062896896\,{x}^{11} \,
+150282885872\,{x}^{10}
\nonumber \\   
\hspace{-0.95in}&&  \, \quad  \quad 
 \,-10690267808\,{x}^{9} \,-6048942832\,{x}^{8} \,
-486602112\,{x}^{7} \,+33772908\,{x}^{6}\,+25075632\,{x}^{5} 
\nonumber \\ 
\hspace{-0.95in}&&  \, \quad  \quad  
 \, +4670454\,{x}^{4}
 \,+13440\,{x}^{3} \,-69066\,{x}^{2} \,-5169\,x \,-63.
\end{eqnarray}

\vskip .1cm 

A degree-34 polynomial $\, P_{34}$ takes place in the
 expression (\ref{RatSolsym2F3}) of the rational solution 
of the symmetric square of the order-three operator $\, F_3$.
This  polynomial $\, P_{34}$ reads:
\begin{eqnarray}
\label{P34}
\hspace{-0.95in}&&  \, \quad 
P_{34}(x)  \, \, \, = \, \, \,17592186044416\,{x}^{34} \,
 -8796093022208\,{x}^{33} \,+204509162766336\,{x}^{32} 
\nonumber \\ 
\hspace{-0.95in}&&  \, \quad \quad 
 \, -240793046482944\,{x}^{31}\,
+347033357516800\,{x}^{30}\,-356447925829632\,{x}^{29}\,
\nonumber \\ 
\hspace{-0.95in}&&  \, \quad \quad 
+307648507412480\,{x}^{28}\, +1547605565767680\,{x}^{27}\,
-1478894410530816\,{x}^{26}\, 
\nonumber \\ 
\hspace{-0.95in}&&  \, \quad \quad 
-3440457380003840\,{x}^{25}\,
+451333349965824\,{x}^{24}\,+3747745613479936\,{x}^{23}\,
\nonumber \\ 
\hspace{-0.95in}&&  \, \quad \quad 
+2236072096432128\,{x}^{22}\,-31693519978496\,{x}^{21}\,
-472806540705792\,{x}^{20}\,
\nonumber \\ 
\hspace{-0.95in}&&  \, \quad \quad 
-202845119840256\,{x}^{19}\,
-55945141010432\,{x}^{18}\, -6522043670528\,{x}^{17}\, 
\nonumber \\ 
\hspace{-0.95in}&&  \, \quad \quad 
+8027346038784\,{x}^{16}\, +5016548481024\,{x}^{15}\,
+1158549638912\,{x}^{14}\,
\nonumber \\ 
\hspace{-0.95in}&&  \, \quad \quad 
 -15663757696\,{x}^{13}\, 
-149163564992\,{x}^{12}\, -59735088608\,{x}^{11}\, -2074333552\,{x}^{10}\,
\nonumber \\ 
\hspace{-0.95in}&&  \, \quad \quad 
 +4173311968\,{x}^{9}\, +738617492\,{x}^{8}\, -85245032\,{x}^{7}\,
-26786428\,{x}^{6}\, +581796\,{x}^{5}\,
\nonumber \\ 
\hspace{-0.95in}&&  \, \quad \quad 
+383308\,{x}^{4}\,-20652\,{x}^{3} \,
 -4867\,{x}^{2}\, +338\,x \, +49.
\end{eqnarray}

\vskip .1cm 

A degree-53 polynomial $\, P_{53}$ takes place in the
 expression (\ref{ratsymadjF3}) of the rational solution 
of the symmetric square of
the adjoint of the (monic) order-three operator $\, F_3$.
This  polynomial $\, P_{53}$ reads:
\begin{eqnarray}
\label{P53}
\hspace{-0.95in}&&  \, \quad 
P_{53}(x)  \, \, \, = \, \, \,
5902958103587056517120\,{x}^{53} \, +4722366482869645213696\,{x}^{52} \,
\nonumber \\ 
\hspace{-0.95in}&&  \, \quad \quad \, \,
 +135675802662133752135680\,{x}^{51} \, 
+36533776637981766975488\,{x}^{50} \,
 \nonumber \\ 
\hspace{-0.95in}&&  \, \quad \quad \, \,
-60743975313220946624512\,{x}^{49} \,
+3954166813899570825658368\,{x}^{48} \, 
\nonumber \\ 
\hspace{-0.95in}&&  \, \quad \quad \, \,
+96486199280696075223040\,{x}^{47} \, -869025933168471881809920\,{x}^{46} \, 
\nonumber \\ 
\hspace{-0.95in}&&  \, \quad \quad \, \,
 +17891408360681834540957696\,{x}^{45} \, 
+19134090460943456531382272\,{x}^{44} \,
\nonumber \\ 
\hspace{-0.95in}&&  \, \quad\quad \, \,
 -6556910656212804697063424\,{x}^{43} \, 
-18888872271338563015016448\,{x}^{42} \, 
 \nonumber \\ 
\hspace{-0.95in}&&  \, \quad \quad \, \,
-9641687070213801940877312\,{x}^{41} \, -856236460709396327956480\,{x}^{40} \, 
\nonumber \\ 
\hspace{-0.95in}&&  \, \quad  \quad \, \,
+1442025047697796450222080\,{x}^{39} +3004016932710650818330624\,{x}^{38} 
\nonumber \\ 
\hspace{-0.95in}&&  \, \quad \quad \, \,
 +2353865809090149001199616\,{x}^{37} \, 
-846024305296182175858688\,{x}^{36} \, 
\nonumber \\ 
\hspace{-0.95in}&&  \, \quad \quad \, \,
-1509872584625178282033152\,{x}^{35} \, 
+296963035049372304801792\,{x}^{34} \, 
\nonumber \\ 
\hspace{-0.95in}&&  \, \quad \quad \, \,
 +832265748859080390213632\,{x}^{33} \, 
+79111183944514552201216\,{x}^{32} \, 
\nonumber \\ 
\hspace{-0.95in}&&  \, \quad \quad \, \,
-245083727451855922397184\,{x}^{31} \, 
-73512832264242582257664\,{x}^{30} \,
 \nonumber \\ 
\hspace{-0.95in}&&  \, \quad \quad \, \,
+31367300451777147568128\,{x}^{29} \, 
+12977767646670109016064\,{x}^{28} \,
 \nonumber \\ 
\hspace{-0.95in}&&  \, \quad \quad \, \,
-3680779152078761099264\,{x}^{27} \,-1138134172734191566848 \,{x}^{26}
\nonumber \\ 
\hspace{-0.95in}&&  \, \quad \quad \, \,
+939259567872233308160\,{x}^{25} \,
 +292487910921964093440\,{x}^{24} \, 
\nonumber \\ 
\hspace{-0.95in}&&  \, \quad \quad \, \,
-129084866249262874624\,{x}^{23} \, 
-72153802925319249920\,{x}^{22}
\nonumber \\ 
\hspace{-0.95in}&&  \, \quad \quad \, \,
+464226011870542848\,{x}^{21}
 \, +8000870954669244416\,{x}^{20} \, 
\nonumber \\ 
\hspace{-0.95in}&&  \, \quad \quad \, \,
+1689242686839294720\,{x}^{19} \, 
-289875180323084800\,{x}^{18} \, 
\nonumber \\ 
\hspace{-0.95in}&&  \, \quad \quad \, \,
 -171427111790469312\,{x}^{17} \,
 -17240190260449408\,{x}^{16} \, +4666400462438480\,{x}^{15} \, 
 \nonumber \\ 
\hspace{-0.95in}&&  \, \quad \quad \, \,
+1816703798900448\,{x}^{14} \,
+258529109814976\,{x}^{13} \,-29106463737504\,{x}^{12}\, 
\nonumber \\ 
\hspace{-0.95in}&&  \, \quad \quad\, \,
-20951763420448\,{x}^{11} 
-2341127444328\,{x}^{10} \, 
+460438019724\,{x}^{9} \,
\nonumber \\ 
\hspace{-0.95in}&&  \, \quad \quad \, \,
+115534150804\,{x}^{8} \, 
-15491040\,{x}^{7} \,
 -1792901976\,{x}^{6} \, -94207344\,{x}^{5}
\nonumber \\ 
\hspace{-0.95in}&&  \, \quad \quad \, \,
+6320658\,{x}^{4}\, -571740\,{x}^{3}
 \, -192705\,{x}^{2} \, -10869\,x \, -147. 
\end{eqnarray}

 \vskip .1cm

\subsection{The polynomial $\, P_{10}$ for $\, \tilde{L}_3$ \newline}
\label{miscelltildeL3}

A degree-ten polynomial $\, P_{10}$ takes place in the
 expression of the rational solution (\ref{ratsymadjtildeL3})  of the 
symmetric square of the adjoint of the (monic) order-three 
operator $\, \tilde{L}_3$.
This  polynomial $\, P_{10}$ reads:
\begin{eqnarray}
\label{P10}
\hspace{-0.95in}&&  \, \quad 
P_{10}(x)  \, \, \, = \, \, \,19394461696\,{x}^{10} \, 
-17411604480\,{x}^{9} \, +6106742784\,{x}^{8} \, 
-1095237312\,{x}^{7}
\nonumber \\ 
\hspace{-0.95in}&&  \, \quad \quad  \quad  \quad 
\, +158668656\,{x}^{6} \,  -36766920\,{x}^{5} \, 
+7627535\,{x}^{4} \, -900594\,{x}^{3}
\nonumber \\ 
\hspace{-0.95in}&&  \, \quad  \quad  \quad  \quad 
 \, +57342\,{x}^{2} \, -1856\,x \, +24.
\end{eqnarray}

\vskip .1cm

\subsection{The polynomials $\, A_{26}(L_4)$  and $\, P_{17}$ for $\, L_4$ \newline}
\label{miscellL4}

The apparent polynomial $\, A_{26}(L_4)$ of the order-four operator $\, L_4$, 
occurs in the expression of the rational solution (\ref{extadjL4})  of the 
exterior square of the adjoint of (the monic order-four operator) 
$\, L_4$, as well as in 
the order-two intertwiner (\ref{homoL4}).
This degree-26 polynomial reads:
\begin{eqnarray}
\label{A26}
\hspace{-0.95in}&&  \, \quad 
A_{26}(L_4)  \, \, \, = \, \, \, \, \, 
 521686412421099571093753036800\,{x}^{26} \,  
\nonumber \\ 
 \hspace{-0.95in}&&  \, \quad  \quad 
-724445324775545659452335063040\,{x}^{25} \, +45081769872830521912080728064\,{x}^{24}
  \nonumber \\ 
 \hspace{-0.95in}&&  \, \quad  \quad 
+616797192523902897669611192320\,{x}^{23} \, -636026962079787427490890252288\,{x}^{22}
 \nonumber \\ 
 \hspace{-0.95in}&&  \, \quad   \quad 
+359505820412663945726355570688\,{x}^{21}\, -142807225508285034141616963584\,{x}^{20} 
  \nonumber \\ 
 \hspace{-0.95in}&&  \, \quad  \quad 
 \, +43345424617004971574289235968\,{x}^{19} \, 
-10332892566359614848157876224\,{x}^{18} 
 \nonumber \\ 
 \hspace{-0.95in}&&  \, \quad  \quad 
+1953967934450852091348254720\,{x}^{17} \, -296338746597146803591135232\,{x}^{16}
 \nonumber \\ 
 \hspace{-0.95in}&&  \, \quad    \quad 
+36761552740911534545901568\,{x}^{15} \, -3850023960384577768909952\,{x}^{14}  
\nonumber \\ 
 \hspace{-0.95in}&&  \, \quad  \quad 
\, +354446803792968575565792\,{x}^{13} \, 
-29645475671183771992224\,{x}^{12} \, 
\nonumber \\ 
 \hspace{-0.95in}&&  \, \quad  \quad 
+2252938824290334087840\,{x}^{11} 
-148605250583921845896\,{x}^{10} \,
\nonumber \\ 
 \hspace{-0.95in}&&  \, \quad  \quad  +7727889974481947660\,{x}^{9} \, 
-251264549473230968\,{x}^{8} \,-1305110830870633\,{x}^{7} 
\nonumber \\ 
 \hspace{-0.95in}&&  \, \quad  \quad 
 +766418384173454\,{x}^{6} 
 \, -52582954690298\,{x}^{5} \, +2099285510560\,{x}^{4}
 \nonumber \\ 
\hspace{-0.95in}&&  \, \quad  \quad 
\, -54037012120\,{x}^{3} \,
 +873083400\,{x}^{2} \, -7854000\,x \, +2800. 
\end{eqnarray}

\vskip .1cm 

A degree-17 polynomial $\, P_{17}$ takes place in the
 expression of the homomorphism (\ref{homL3}) of $\,\tilde{L}_3$
with its adjoint, as well as in the rational solution
(\ref{RatSolext2L4}) of the exterior square of $\, L_4$ and the 
rational solution (\ref{extadjL4}) of the 
exterior square of the adjoint of (the monic order-four operator) 
$\, L_4$. This  polynomial $\, P_{17}$ reads:
\begin{eqnarray}
\label{P17}
\hspace{-0.95in}&&  \, \quad 
P_{17}(x)  \, \, \, = \, \, \,
140082179425173504\,{x}^{17} \, 
-496507256028790784\,{x}^{16} \,
 \nonumber \\ 
\hspace{-0.95in}&&  \, \quad  \quad  \quad 
 +705909942330064896\,{x}^{15} \,
-440315308230574080\,{x}^{14} \,
\nonumber \\ 
\hspace{-0.95in}&&  \, \quad \quad  \quad 
 +141123001405931520\,{x}^{13} \, 
-25595376023494656\,{x}^{12} 
 \, +4059589860750336\,{x}^{11} \, 
\nonumber \\ 
\hspace{-0.95in}&&  \, \quad \quad  \quad 
-1133589089074624\,{x}^{10} \, 
+350453101085400\,{x}^{9} \,
 -74115473257440\,{x}^{8} \,
\nonumber \\ 
\hspace{-0.95in}&&  \, \quad \quad  \quad
+10126459925120\,{x}^{7} \, 
 -904049598675\,{x}^{6} \, +52738591890\,{x}^{5} 
 \, -1959091320\,{x}^{4} \,
\nonumber \\ 
\hspace{-0.95in}&&  \, \quad \quad  \quad
+43407720\,{x}^{3} \, -502593\,{x}^{2} \, +2548\,x \, -12. 
\end{eqnarray}

\vskip .1cm

\section{Some linear differential operators appearing in Section 3}
\label{appendixSomeLn}

\subsection{Linear differential operators for $\, F_3$}
\label{appendixSomeLnF3}

The order-two differential operator $\, R_2$ occurring
 in the solution of $\, F_3$ given in (\ref{solF3}) reads
\begin{eqnarray}
\hspace{-0.95in}&&  \quad \quad  \quad   R_2 \, \,=\,\,\, 
  -{\frac{x \cdot \,  (1-4x^2)(1-16x^2)
 \cdot \,  P_{15}(x)}{D_5(x) \cdot \, A_7(F_2)}} \cdot \,D_x^2\,  
\nonumber \\
\hspace{-0.95in}&&  \qquad \qquad \qquad  \qquad 
 - {\frac{P_{20}(x)}{D_5(x) \cdot \, A_7(F_2)}} \cdot \,D_x
\,\,  \, 
+ {\frac{8 x \cdot \,P_{22}(x)}{D_5(x) \cdot \, A_7(F_2)}}, 
 \nonumber
\end{eqnarray}
where $\, A_7(F_2)$ is the apparent polynomial of $\, F_2$ given in (\ref{appA7F2}),
and where: 
\begin{eqnarray}
\hspace{-0.95in}&& \quad   \quad
 D_5(x) \,\,= \,\,\, \, x \cdot \, (1-x)  \, (1+2\,x)  \, (4\,{x}^{2}+3\,x+1),
 \nonumber \\
\hspace{-0.95in}&& \quad   \quad 
P_{15}(x)\,  \,=\, \, \,\, 
 2\, +9\,x\, -99\,{x}^{2}\, -873\,{x}^{3} \, -1865\,{x}^{4} \, +12140\,{x}^{5}\, +83412\,{x}^{6}
  \nonumber \\
\hspace{-0.95in}&& \quad   \qquad \quad
 +238912\,{x}^{7}\, +375008\,{x}^{8}\, -1397504\,{x}^{9}\, -9548288\,{x}^{10}\, -17188864\,{x}^{11}  
\nonumber \\
\hspace{-0.95in}&& \quad   \qquad \quad
 -7581696\,{x}^{12}\, +2260992\,{x}^{13}\, -7471104\,{x}^{14}\, -786432\,{x}^{15},
 \nonumber \\
\hspace{-0.95in}&& \quad   \quad 
P_{22}(x) \,\, =\,\, \,
2\,\, -5\,x\, \,-233\,{x}^{2}\,\, +43\,{x}^{3}\, +7343\,{x}^{4}\, +14408\,{x}^{5}\, +28660\,{x}^{6} 
 \nonumber \\
\hspace{-0.95in}&& \quad   \quad \qquad
 +68224\,{x}^{7}\, -2196448\,{x}^{8} \, -13292608\,{x}^{9} \,
 -21440000\,{x}^{10}\, +94341632\,{x}^{11}
  \nonumber \\
\hspace{-0.95in}&& \quad   \quad \qquad  
 +562065408\,{x}^{12}\, +700620800\,{x}^{13}\, 
-1591803904\,{x}^{14}\, -4451794944\,{x}^{15}
 \nonumber \\
\hspace{-0.95in}&& \quad   \quad \qquad
 -2017984512\,{x}^{16}\, +467664896\,{x}^{17} \, -2013265920\,{x}^{18}\,  -268435456\,{x}^{19}
 \nonumber \\
\hspace{-0.95in}&& \quad   \quad \quad \quad \, \,  
+67108864\,{x}^{20},
 \nonumber \\
\hspace{-0.95in}&& \quad   \quad 
P_{18}(x)\,  \,=\, \, \, 
4\,\, +13\,x\,\, +87\,{x}^{2}\, +1472\,{x}^{3}\, -1950\,{x}^{4}\, -34896\,{x}^{5}\, +53220\,{x}^{6} 
 \nonumber \\
\hspace{-0.95in}&& \quad   \quad \qquad +630696\,{x}^{7}\, +536416\,{x}^{8}\,
 -4436416\,{x}^{9}\, -21416192\,{x}^{10}\, -32954368\,{x}^{11} 
 \nonumber \\
\hspace{-0.95in}&& \quad   \quad \qquad +80510976\,{x}^{12}\, +304701440\,{x}^{13}
 \, +227115008\,{x}^{14}\, +5636096\,{x}^{15}
 \nonumber \\
\hspace{-0.95in}&& \quad   \quad \qquad
 +151519232\,{x}^{16}\, +29360128\,{x}^{17}\, -16777216\,{x}^{18}.
  \nonumber
\end{eqnarray}

\vskip .1cm

\subsection{Linear differential operators for $\, \tilde{L}_3$}
\label{appendixSomeLntildeL3}

The order-two linear differential operator $\, R_2$ occurring
 in the solution of $\, \tilde{L}_3$ given in (\ref{soltL3}) is
\begin{eqnarray}
\hspace{-0.95in}&& \quad  \quad  \quad  
 R_2 \,\, \,  =  \, \, \,  \, 
{\frac {{x}^{2} \cdot \, Q_3(x)\cdot \,  P_4(x) }
{ (1-4\,x)  \, (1-16\,x)^{7}}} \cdot \, D_x^2 \,  \,  \, 
+{\frac {x \cdot \, Q_3(x) P_6(x) }{ (1-4\,x)^{2} \, (1-16\,x)^{8}}} \cdot \, D_x
 \nonumber \\
\hspace{-0.95in}&& \quad  \qquad \qquad  \qquad \quad  \quad  
 +{\frac {216\, {x}^{3} \cdot \, (1-12\,x) \cdot \, Q_3(x) \cdot \,  P_3(x) }{ (1-4\,x)^{3}
 \, (1-16\,x)^{9}}}, 
\end{eqnarray}
with:
\begin{eqnarray}
\hspace{-0.95in}&& \quad   \quad
 Q_3 \,\, = \,\, 5\, -160\,x\,+1232\,{x}^{2}\,-1024\,{x}^{3},
 \nonumber \\
\hspace{-0.95in}&& \quad  \quad 
P_4 \,\, = \,\,\, 
3\,-68\,x\,+976\,{x}^{2}\,-2624\,{x}^{3}\,-61440\,{x}^{4},
 \nonumber \\
\hspace{-0.95in}&& \quad   \quad
 P_6 \,\, = \,\,\, 
3\,-152\,x\,+3528\,{x}^{2}\,-42384\,{x}^{3}\,
+89024\,{x}^{4}\,+1966080\,{x}^{5}\,-5898240\,{x}^{6}, 
\nonumber \\
\hspace{-0.95in}&& \quad   \quad 
P_3 \,\,= \,\,\, 3\,-92\,x\,+1792\,{x}^{2}\,+4096\,{x}^{3}.
 \nonumber
\end{eqnarray}

\vskip .1cm 

\subsection{The linear differential operator  $\, L_4$}
\label{appendixSomeLnL4}

The linear differential operator $\, L_4$ has been analyzed
in~\cite{2011-calabi-yau-ising} and was shown to be 
equivalent to a {\em Calabi-Yau equation} with solution
\begin{eqnarray}
\label{4F3good}
\hspace{-0.95in}&& \quad  \quad \qquad \qquad 
_4F_3\Bigl( [{{1} \over {2}},\, {{1} \over {2}},\,
 {{1} \over {2}},\, {{1} \over {2}}], [1,1,1]; \, \, z  \Bigr),  
\end{eqnarray}
where the argument $\, z$ is an {\em algebraic pullback}, in 
the variable 
$ \,w = \, s/2/(1+s^2)$,
(with $s= \, \sinh(2K)$, where $\, K= \, J/kT$ is 
the Ising model coupling constant):
\begin{eqnarray}
\label{pullsquare}
\hspace{-0.95in}&& \quad \quad \quad \qquad 
z \, \,\,  = \, \,\,\,  
 \Bigl({{ 1\, + \,  (1\, -16 \cdot  w^2)^{1/2}} 
\over { 1 \, - \,  (1\, -16 \cdot  w^2)^{1/2}}}  \Bigr)^4
 \,\,\,  = \, \,\,\,  s^8.
\end{eqnarray}
Note that the variable $\, w$ deals equally with the high and low regime
of temperature. One has another pullback which is $ \, 1/z= \, 1/s^8$.

\vskip .1cm
\subsubsection{Direct sum structure associated with $\, L_4$ \newline}
\label{subappendixSomeLnL4}

 When written in the variable $\, x= \, w^2$,
the $ \, _4F_3$ hypergeometric function 
{\em with any of the two pullbacks}, 
for instance the series with {\em integer coefficients}
\begin{eqnarray}
\label{simple4F3}
\hspace{-0.95in}&& \quad 
_4F_3\Bigl( [{{1} \over {2}},\, {{1} \over {2}},\,
 {{1} \over {2}},\, {{1} \over {2}}], [1,1,1]; \, \, {{1} \over {z}}  \Bigr)
\,\,  = \, \,\,\,  1 \, +16\,{x}^{4}\,+512\,{x}^{5}\,+11264\,{x}^{6}\,
\, + \,\,\cdots, 
\end{eqnarray} 
 is annihilated by an 
{\em order-eight} linear operator
$\, L_8 \, = \, \, H_4^{(1)} \oplus \, H_4^{(2)}$
which is a {\em direct sum} 
of {\em two order-four} linear differential operators $\, H_4^{(1)}$ 
and $ \, H_4^{(2)}$:
\begin{eqnarray}
\hspace{-0.95in}&& \quad    H_4^{(1)} \, \,  \, =\, \,\,  \, 
 {x}^{3} \cdot \, (1-16\,x)^{2} \, (1-8\,x)^{4} \cdot \, D_x^4 
\nonumber \\
\hspace{-0.95in}&& \quad   \quad \quad \quad 
+ 2\,{x}^{2} \cdot \, (1-8\,x)^{3} \, (1-16\,x)
  \, (512\,{x}^{2} \, -96\,x \, +3) \cdot   \, D_x^3 
\nonumber \\
\hspace{-0.95in}&& \quad   \quad \quad \quad 
+ x \cdot \,  (1-8\,x)^{2} \, (233472\,{x}^{4} \, 
-83968\,{x}^{3}+10880\,{x}^{2}-512\,x+7) 
\cdot  \, D_x^2
 \nonumber \\
\hspace{-0.95in}&& \quad   \quad \quad \quad 
 - \, (1-8\,x)  \, (589824\,{x}^{5} \, -266240\,{x}^{4} \, 
+40960 \,{x}^{3}-3968\,{x}^{2}+144\,x-1) \cdot \, D_x
 \nonumber \\
\hspace{-0.95in}&& \quad   \quad \quad \quad 
  -256\, x \cdot \, (1 -16\,x -128\,{x}^{2}), 
 \nonumber
\end{eqnarray}
\begin{eqnarray}
\hspace{-0.95in}&& \quad   H_4^{(2)} \,\, \, = \,\,\,\, 
 {x}^{3} \cdot \, (1-16\,x)^{4} \, (1-8\,x) \cdot   \, D_x^4 
\nonumber \\
\hspace{-0.95in}&& \quad   \quad \quad  \qquad 
 + 2\,{x}^{2} \cdot \, (1-16\,x)^{3} \, 
 \left( 640\,{x}^{2}-96\,x+3 \right) \cdot  \, D_x^3 
\nonumber \\
\hspace{-0.95in}&& \quad   \quad \quad \qquad 
 - x \cdot \, (1 -16\,x)^{2}
 \, (50688\,{x}^{3}-8768\,{x}^{2}+456\,x-7)  \cdot   \, D_x^2
 \nonumber \\
\hspace{-0.95in}&& \quad   \quad \quad \qquad  + \, (1-16\,x) 
 \, (466944\,{x}^{4}\, -90112\,{x}^{3} \, +5952\,{x}^{2} \, -144\,x \, +1) \cdot \, D_x
 \nonumber \\
\hspace{-0.95in}&& \quad   \quad \quad \qquad 
+ 256\, x \cdot  \, (1-8\,x) 
\, (192\,{x}^{2} \, -16\,x \, +1).  
 \nonumber
\end{eqnarray}
Each one corresponds to a 
Calabi-Yau ODE~\cite{2011-calabi-yau-ising}.

Note that these two order-four operators are simply conjugated:
\begin{eqnarray}
\label{conjH1H2}
\hspace{-0.95in}&& \quad  \qquad  \qquad 
{\frac {\sqrt {1-16\,x}}{1-8\,x}} \cdot \,  H_4^{(1)}
 \,\,  \, = \,\,\,\,  \, H_4^{(2)} \cdot \, {\frac {\sqrt {1-16\,x}}{1-8\,x}}. 
\end{eqnarray}

The solution of $\, H_4^{(2)}$ analytic at $\, x \, = \, \, 0$ 
is the series with {\em integer coefficients}:
\begin{eqnarray}
\label{solH42}
\hspace{-0.95in}&& \quad 
sol(H_4^{(2)}) \,\,\,  = \,\,\,\, {{1} \over {4}} \cdot \, 
 {\frac{\sqrt{1-16 x}}{x \cdot \, z^{1/4}}} \cdot \,  \, 
_4F_3\Bigl( [{{1} \over {2}},\, {{1} \over {2}},\,
 {{1} \over {2}},\, {{1} \over {2}}], [1,1,1]; \, \, {1 \over z}  \Bigr)
\\
\hspace{-0.95in}&& \quad  \quad  \quad    \quad  
  = \,\,\,\, 1 \, -16\,{x}^{2} \, -256\,{x}^{3} \, -3568\,{x}^{4} \, 
-48640\,{x}^{5} \, -664832\,{x}^{6}   \,  + \,  \, \cdots 
\nonumber 
\end{eqnarray} 
The solution of $\, H_4^{(1)}$ analytic at $\, x \, = \, \, 0$ 
is the series with {\em integer coefficients}:
\begin{eqnarray}
\label{solH41}
\hspace{-0.95in}&& \quad 
sol(H_4^{(1)}) \,\,\,  = \,\,\,\, 
1 \, +16\,{x}^{2} \, +256\,{x}^{3}\, +3600\,{x}^{4}\,
+49664\,{x}^{5}\,+687360\,{x}^{6}
\nonumber \\
\hspace{-0.95in}&& \quad \quad \quad \quad   \quad  
\,+9596928\,{x}^{7}\, +135300368\,{x}^{8}\, +1925268480\,{x}^{9} 
\, \,  + \,\, \cdots
\end{eqnarray} 
The simple $\, _4F_3$ hypergeometric function (\ref{simple4F3}) is actually equal to 
 the half sum $\, (sol(H_4^{(1)}) +sol(H_4^{(2)}))/2$,  of 
the two solutions (\ref{solH42}) and (\ref{solH41})
of $\, H_4^{(2)}$ and $\, H_4^{(1)}$.

\vskip 0.1cm

\subsubsection{Solution of the linear differential operators for $\, L_4$ \newline}
\label{subappendixSomeLnL4}

The order-four operator $\, H_4^{(2)}$ 
is {\em homomophic to the order-four operator} $\, L_4$,
emerging as a factor operator for $\, \tilde{\chi}^{(6)}$:
\begin{eqnarray}
\label{homL4H2}
\hspace{-0.95in}&& \quad \, \, \qquad \qquad  \qquad  
S_3  \cdot \,  H_4^{(2)} \,\,\,  = \,\,\,\, L_4 \cdot \, R_3, 
\end{eqnarray}
where $\, S_3$ and $\, R_3$ are two order-three intertwiners.
One immediately deduces the solution of $\, L_4$ given 
in terms of the intertwiner $\, R_3$ in (\ref{homL4H2})
acting on the solution of the order-four operator $\, H_4^{(2)}$:
\begin{eqnarray}
\label{R3insolL4}
\hspace{-0.95in}&& \quad \qquad \qquad \qquad  
sol(L_4) \, \,\, =\,\,\,\,  R_3 \Bigl(sol(H_4^{(2)})\Bigr),  
\end{eqnarray}
where the order-three linear differential operator
 $\,R_3$ reads:
\begin{eqnarray}
\hspace{-0.95in}&& \quad \quad  \quad 
36\,{x}^{5}\cdot \, (1-4\,x) 
 \, (1-16\,x)^{11} \cdot \, A_4(\tilde{L}_3) \cdot \, R_3
\nonumber \\
\hspace{-0.95in}&& \quad \quad \quad  \quad
 \, \, \,\, \, \, \,=\,\,\,
-{x}^{2} \cdot \, (1-8\,x) 
 \, (1-16\,x)^{6} \cdot \, Q_3 \cdot \,D_x^{3}\, \, 
  -x \left( 1-16\,x \right)^{5}\cdot \, Q_2 \,\cdot \,  D_x^{2}
 \nonumber \\
\hspace{-0.95in}&& \quad  \quad \quad \qquad \quad \qquad 
\, - \, (1-16\,x)^{4} \cdot \, Q_1 \,\cdot \, D_x \, \, 
-128\,x \cdot \, (1-16\,x)^{3} \cdot \, Q_0, 
\end{eqnarray}
\begin{eqnarray}
\hspace{-0.95in}&& \quad   \quad 
Q_3\,  \,=\, \, \, 
20-2270\,x+106086\,{x}^{2}-2675757\,{x}^{3} +40471555\,{x}^{4} -389549218\,{x}^{5}
 \nonumber \\
\hspace{-0.95in}&& \quad   \qquad \quad +2566958582\,{x}^{6}-13288554644\,{x}^{7}
 \, +53910201600\,{x}^{8}-95886464512\,{x}^{9}
 \nonumber \\
\hspace{-0.95in}&& \quad   \qquad \quad -40752267264\,{x}^{10}
 +93413441536\,{x}^{11} +82141249536\,{x}^{12}, 
 \nonumber \\
\hspace{-0.95in}&& \quad   \quad 
Q_2\,  \,=\, \, \, 
60-8730\,x+551602\,{x}^{2}-19952295\,{x}^{3} +459567769\,{x}^{4} -7113445902\,{x}^{5}
 \nonumber \\
\hspace{-0.95in}&& \quad   \qquad \quad +76621809730\,{x}^{6} 
-596173812436\,{x}^{7} +3524748623424\,{x}^{8}
 \nonumber \\
\hspace{-0.95in}&& \quad  \qquad \quad -16119878544384\,{x}^{9} \, 
+49591145041920\,{x}^{10} \, -62942370168832\,{x}^{11} 
\nonumber \\
\hspace{-0.95in}&& \quad  \qquad \quad -43186282037248\,{x}^{12}
+62103616487424\,{x}^{13} \, +63084479643648\,{x}^{14},
 \nonumber \\
\hspace{-0.95in}&& \quad   \quad 
Q_1\,  \,=\,\, \,  
20\,-3710\,x\,+303254\,{x}^{2}-14374525\,{x}^{3} +439222171\,{x}^{4} -9126353218\,{x}^{5}
 \nonumber \\
\hspace{-0.95in}&& \quad   \qquad \quad +133114097446\,{x}^{6} -1396508587356\,{x}^{7} 
+10831258373280\,{x}^{8}
 \nonumber \\
\hspace{-0.95in}&& \quad   \qquad \quad -63997739175680\,{x}^{9}
+285429913462784\,{x}^{10}-832850214682624\,{x}^{11}
 \nonumber \\
\hspace{-0.95in}&& \quad   \qquad \quad +969294168981504\,{x}^{12} \, 
+842807128358912\,{x}^{13}-1089827550265344\,{x}^{14}
 \nonumber \\
\hspace{-0.95in}&& \quad  \qquad \quad -1135520633585664\,{x}^{15},
 \nonumber \\
\hspace{-0.95in}&& \quad   \quad 
Q_0 \, \,=\, \, \, 
20-2590\,x+145574\,{x}^{2}-4725757\,{x}^{3}+99952043\,{x}^{4} -1473719054\,{x}^{5}
 \nonumber \\
\hspace{-0.95in}&& \quad   \qquad \quad +15848325886\,{x}^{6}-128583477160\,{x}^{7} \, 
+795236207808\,{x}^{8}-3570673925376\,{x}^{9}
 \nonumber \\
\hspace{-0.95in}&& \quad   \qquad \quad +9940600639488\,{x}^{10} 
\, -10105313820672\,{x}^{11} \, 
-13061917245440\,{x}^{12} 
\nonumber \\
\hspace{-0.95in}&& \quad  \qquad \quad
 +14868774125568\,{x}^{13}+15771119910912\,{x}^{14}, 
 \nonumber
\end{eqnarray}
the apparent polynomial $\, A_4(\tilde{L}_3)$ being given in (\ref{A4tildeL3}). 

\section{Homomorphism with the adjoint for order-three and order-four operators}
\label{genericextsym}

We show here, starting with {\em generic} (and irreducible) operators,
 the link between the homomorphism with the adjoint
and the occurrence of a rational solution of the symmetric (or exterior) square of the
differential operator for operators of order three and four.

For a linear differential operator of order $\, q$, the order 
of the intertwiner in an equivalence 
relation may reach the order $\, q\, -1$.
\ref{genericL3R210} considers order-three generic operator with intertwiner of order
two, one and zero (i.e. a function).
\ref{genericL4R20} deals with order-four generic operator with order-two and order-zero
intertwiner, and \ref{genericL4R31} is for the case of 
order-three and order-one intertwiner.

\subsection{Order-three linear differential operator}
\label{genericL3R210}

With the generic order-three differential operator $\,L_3$
\begin{eqnarray}
\hspace{-0.95in}&& \quad  \qquad  \quad  
L_3 \,\,\, =\,\,\,\, 
D_x^3 \, \,\,  + p_2(x) \cdot \,D_x^2\,\,\,  + p_1(x) \cdot \, D_x\,\, \,  + p_0(x), 
\end{eqnarray}
and the order-two differential operator
\begin{eqnarray}
\label{theR2}
\hspace{-0.95in}&& \quad  \qquad  \quad  
R_2\,\, \,=\,\,\, \, a_2(x) \cdot \, D_x^2\,\, \,+ a_1(x) \cdot \,D_x \,\,\, + a_0(x), 
\end{eqnarray}
one demands that the relation
\begin{eqnarray}
\label{relationHomoL3}
\hspace{-0.95in}&& \quad   \qquad  \quad  
 L_3 \cdot \,R_2 \,\,\, = \,\,\,\,  adjoint(R_2) \cdot\, adjoint(L_3), 
\end{eqnarray}
be fulfilled, which means that $\, L_3$ is homomorphic to its adjoint.

Zeroing the expressions in front of each derivative $\, D_x^j$
 in (\ref{relationHomoL3})
gives a set of equations which solve as
\begin{eqnarray}
\label{thea2x}
\hspace{-0.95in}&& \quad   \quad   \quad    \quad  
a_2(x) \,\,=\,\, \,\, sol \left( Sym^2(L_3) \right)
\end{eqnarray}
\begin{eqnarray}
\label{thea1x}
\hspace{-0.95in}&& \quad   \quad    \quad    \quad  
a_1(x)\, \,=\, \,\,
- p_2(x) \cdot \, a_2(x)\, \, 
 -{1 \over 2} \, {\frac {d a_2(x)}{dx}}, 
\end{eqnarray}
and
\begin{eqnarray}
\label{thea0x}
\hspace{-0.95in}&& \quad   \quad    \quad    \quad  
a_0(x)\, \,\, =\,\,\, \,\,  N_5 \cdot \,a_2(x).
\end{eqnarray}
The order-five differential operator $\, N_5$ is such that
\begin{eqnarray}
\hspace{-0.95in}&& \,  C_1^{(3)} \cdot \, N_5 \,\,=\,\,\,
9 \,D_x^{5} \, 
+30\,p_2(x) \cdot \, D_x^{4} \,+ Q_3 \cdot \,D_x^{3}
\,+ Q_2 \cdot \,D_x^{2} \,+ Q_1 \cdot \,D_x \, + \, Q_0, 
\end{eqnarray}
where
\begin{eqnarray}
\hspace{-0.95in}&& \quad  \quad Q_3 \,\,=\,\,\,\,
25\,  p_2(x)^{2} \,\,   +45\,p_1(x) \, \,  
+15\,{\frac {d p_2(x)}{dx}},
 \nonumber \\
\hspace{-0.95in}&& \quad   \quad 
Q_2 \,\,=\,\,\,\,
75\,p_1(x)\, p_2(x)\,\,   
 +45\,p_0(x)\, \,   +45\,{\frac {d p_1(x)}{dx}},
 \nonumber \\
\hspace{-0.95in}&& \quad   \quad 
Q_1\, \,=\,\,\,\,
36\, p_1(x)^{2}\,\,   -4\, p_2(x)^{4} \, \,  
+42\,p_2(x)\,  p_0(x) \,  
+22\, p_2(x)^{2}\,   p_1(x)
 \nonumber \\
\hspace{-0.95in}&& \quad   \qquad \qquad 
+9\,{\frac {d^{2} p_1(x)}{d{x}^{2}}}\, \,  
 +63\,{\frac {d p_0(x)}{dx}} \,\,  
-18\, {\frac {d p_2(x)}{dx}}\,    p_2(x)^{2}\,  
+48\,p_2(x)  \, {\frac {d}{dx}} p_1(x) 
 \nonumber \\
\hspace{-0.95in}&& \quad   \qquad \qquad 
-3\,p_1(x)  \, {\frac {d p_2(x)}{dx}} \, \,  
  -9\, p_2(x) \,  {\frac {d^{2} p_2(x)}{d{x}^{2}}},
 \nonumber \\
\hspace{-0.95in}&& \quad   \quad 
Q_0 \,\, =\,\,\,
36\, p_1(x)^{2} p_2(x)\, \,  \,  
 -36\, p_1(x) \,  p_0(x) \,  \,  
-8\, p_1(x)  \, p_2(x)^{3} \, \,  
+8\, p_2(x)^{2} p_0(x)\,  
 \nonumber \\
\hspace{-0.95in}&& \quad   \quad \qquad 
-18\, {\frac {d^{2} p_2(x) }{d{x}^{2}}} \,  p_1(x) \,   \, 
+18\, {\frac {d^{2} p_2(x) }{d{x}^{2}}}  {\frac {d p_2(x)}{dx}}  \,   \, 
+36\, ({\frac {d}{dx}} \,  p_2(x))^{2}  \, p_2(x)\,  
 \nonumber \\
\hspace{-0.95in}&& \quad   \quad \qquad 
+102\,  {\frac {d p_2(x)}{dx}}\,   p_0(x) \,   \, 
-54\, {\frac {d p_2(x)}{dx}} \,   {\frac {d p_1(x)}{dx}} \,    \, 
+8\, {\frac {d p_2(x)}{dx}}\,   p_2(x)^{3}\, 
 +18\,{\frac {d^{2} p_0(x)}{d{x}^{2}}}
 \nonumber \\
\hspace{-0.95in}&& \quad   \quad \qquad 
+42\, p_2(x) \, \, {\frac {d p_0(x) }{dx}}  \,  \, 
+54\, p_1(x)  \,{\frac {d p_1(x)}{dx}}     \,  \, 
-72\, {\frac {d p_2(x)}{dx}} \, \, p_1(x) \,   p_2(x), 
 \nonumber 
\end{eqnarray}
and
\begin{eqnarray}
\hspace{-0.95in}&& \quad   \quad 
C_1^{(3)} \,\, \, =\,\, \, \,
36\, p_1(x) \, p_2(x) \,\, \,   -108\, p_0(x)  \, \,   \,
 -8\, p_2(x)^{3}  \, \,  \, 
 -36\, p_2(x) \,\,   {\frac {d p_2 \left( x \right)}{dx}} 
\nonumber \\
\hspace{-0.95in}&& \quad  \qquad \qquad \qquad  \, \,
+54\,{\frac {d  p_1(x)}{dx}} \, \,  
-18\,{\frac {d^{2} p_2(x)}{d{x}^{2}}}.
 \nonumber
\end{eqnarray}
The last expression, when $\,C_1^{(3)}\,= \,\, 0$, is 
the condition for $ \, L_3$ to have an order-five
symmetric square instead of the order six.

From the expression (\ref{thea2x}), one sees that if $\,Sym^2(L_3)$ has a 
{\em rational solution}, one may take $\,a_2(x)$ {\em as this solution}, 
and automatically the expressions of $a_1(x)$ and $\,a_0(x)$ will be rational. 
If $\,Sym^2(L_3)$ has no rational solution (assume $\,Sym^2(L_3)$ is irreducible), 
one may still take for $\,a_2(x)$ {\em any solution} of $\, Sym^2(L_3)$ 
and with the corresponding
$\,a_1(x)$, $\,a_0(x)$, the relation (\ref{relationHomoL3}) will be verified.
But now, the intertwiner $\,R_2$ is {\em no more over the rationals}.

Note that, one may use (\ref{thea2x}) and (\ref{thea0x}) with its derivative,
to reduce (\ref{thea0x}) to the following inhomogeneous differential equation
\begin{eqnarray}
\label{otherthea0x}
\hspace{-0.95in}&& \quad   \quad  \quad   \quad  
6\, {\frac {d a_0(x)}{dx}}\, \, +  4\, p_2(x) \cdot \,a_0(x)
\,\,  \, =\,\,\, \,
E_3 \cdot \, a_2(x)
\end{eqnarray}
where the order-three differential operator $\, E_3$ is
\begin{eqnarray}
\hspace{-0.95in}&& \quad  \quad  E_3 \,\,\, = \,\,\,\,\,
D_x^{3}\,\,\, +3 \,p_2(x)\cdot \, D_x^{2} \,\,\,
+ \, \Bigl( 4\, p_1(x)\, +2\,  p_2(x)^{2}
 -3\,{\frac {d p_2x)}{dx}}  \Bigr) \cdot \, D_x 
\nonumber \\
\hspace{-0.95in}&& \quad   \qquad \quad \,\,
+4\,p_1(x)\,  p_2(x)\, \, -4\,p_0(x)\,\,
-6\,{\frac {d^{2} p_2(x)}{d{x}^{2}}} \, \,
 +6\,{\frac {d p_1(x)}{dx}} \,  \,
 -4\,p_2(x) \, {\frac {d p_2(x)}{dx}}. 
 \nonumber
\end{eqnarray}

The way to obtain $\,a_0(x)$, via (\ref{thea0x}), is 
more tractable, since this amounts
to taking derivatives of the rational solution $\, a_2(x)$.
The route via (\ref{otherthea0x}) calls
 for an integration, and re-injection in 
(\ref{relationHomoL3}), to fix the constants of integration.

\vskip 0.2cm

Instead of an intertwiner $\, R_2$ of order two, let us consider the situation
with an order-one intertwiner, $ \, a_1(x) \cdot \,D_x \, + a_0(x)$. In 
this case, one obtains 
\begin{eqnarray}
\hspace{-0.95in}&& \quad  
a_0(x)\,  \,=\, \, \,  \, 
- p_2(x) \cdot \, a_1(x)\, \,  -{\frac {d a_1(x)}{dx}}
 \quad \,  \hbox{and:} \,  \quad \quad
 a_1(x)\,  \,=\,\,\,   sol(Ext^2(L_3)).
\end{eqnarray}
Recall that (with $\,W_L(x)$ the Wronskian of $\,L_3$)
\begin{eqnarray}
\hspace{-0.95in}&& \quad \quad  \quad  \quad 
Ext^2(L_3) \cdot \, W_L(x)\,\, \,   =\, \, \,  \, W_L(x) \cdot adjoint(L_3).
\end{eqnarray}
Since $\, L_3$ is irreducible, $\, a_1(x)$ cannot be rational.
Therefore, there is no homomorphism between $\, L_3$ and its adjoint
with an order-one intertwiner over the rationals.
For order zero intertwiner, i.e. a function $a_0(x)$, one obtains
\begin{eqnarray}
\hspace{-0.95in}&& \quad \quad \quad \qquad  
 a_0(x) \, \,=\,\,\,   W_L(x)^{2/3}\, 
 \qquad {\rm and} \, 
 \qquad \quad  C_1^{(3)} \, \,=\,\,\,  0, 
\end{eqnarray}
where $\, W_L(x)$ is the Wronskian of $\, L_3$.
The condition $\, C_1^{(3)} \,=\,0$ makes the symmetric square of $\, L_3$ 
of order five, and $\, L_3$ is the
 symmetric square of an order-two differential operator.

\subsection{Order-four linear differential operator}
\label{genericL4R20}

What we have done for the generic order-three linear  differential operator
can be repeated for a generic order-four operator $L_4$.

For the order-four differential operator
\begin{eqnarray}
\hspace{-0.95in}&& \quad \quad \,  \quad    
L_4 \,\,\,  = \,\,\, \, \, 
D_x^4 \,\,  \,+p_3(x)\cdot \,  D_x^3 \, \,\,  + p_2(x)\cdot \,  D_x^2 \,  \,
\, + p_1(x) \cdot \,  D_x\, \,   + p_0(x), 
\end{eqnarray}
and an order-two operator as in (\ref{theR2}), the relation 
\begin{eqnarray}
\label{relationHomoL4}
\hspace{-0.95in}&& \quad  \quad \quad \quad    
 L_4 \cdot \,  R_2 \,\,\,\, =\,\,\,\, \,
adjoint(R_2) \cdot \, adjoint(L_4), 
\end{eqnarray}
is solved to give
\begin{eqnarray}
\label{a2extorderfour}
\hspace{-0.95in}&& \quad  
a_2(x) \,\,=\, \,\, sol(Ext^2(L_4)),
  \quad  \quad 
 a_1(x) \, \,= \, \,\,
- p_3(x) \cdot \, a_2(x)\, \,\, -{\frac {d a_2(x)}{dx}}, 
\end{eqnarray}
and
\begin{eqnarray}
\label{thea0xE}
a_0(x)\, \,=\,\,\, \, N_5 \cdot \, a_2(x).
\end{eqnarray}

The order-five differential operator $ \, N_5$ is such that
\begin{eqnarray}
\hspace{-0.95in}&& \,
C_1^{(4)} \cdot \, N_5 \,\,  =\, \, \, 
4\,D_x^{5} \, 
+10\,p_3(x) \cdot \,  D_x^{4} \, 
+Q_3 \cdot \,D_x^{3} \, 
+Q_2 \cdot \,D_x^{2} \,  \, +Q_1 \cdot \,D_x \, +Q_0,
\end{eqnarray}
where
\begin{eqnarray}
\hspace{-0.95in}&& \quad   \quad 
Q_3 \, \,=\,\,  \, \,
7\, (p_3(x))^{2} \, \, +8\,p_2(x) \,  \, 
+8\,{\frac {d p_3(x)}{dx}},
 \nonumber \\
\hspace{-0.95in}&& \quad   \quad 
Q_2 \, \, = \,\,  \, \,
14\,p_2(x) \, p_3(x)\,  \, 
 -4\,p_1(x)  \, 
+16\,{\frac {d p_2(x)}{dx}},
 \nonumber \\
\hspace{-0.95in}&& \quad   \quad 
Q_1 \,\, = \,\,\,\,
4\, p_2(x)^{2}\, \,  -16\, p_0(x) \,  \,  \, 
+5\, p_3(x)^{2} \,  p_2(x) \,  \, 
-2\, p_3(x) \,  p_1(x) \,  \,  - p_3(x)^{4} 
 \nonumber \\
\hspace{-0.95in}&& \quad   \qquad \qquad 
 +8\,{\frac {d p_1(x)}{dx}}  \,  \, 
+4\,{\frac {d^{2} p_2(x)}{d{x}^{2}}}  \,  \,  
 -6\, p_3(x)^{2}{\frac {d p_3(x)}{dx}}   \,  \, 
  +14\, p_3(x) \, {\frac {d p_2(x)}{dx}}
 \nonumber \\
\hspace{-0.95in}&& \quad   \qquad \qquad 
-4\, p_3(x) {\frac {d^{2} p_3(x)}{d{x}^{2}}},
 \nonumber \\
\hspace{-0.95in}&& \quad   \quad
 Q_0 \,\,= \, \, \,
4\, p_3(x) \,  p_2(x)^{2} \,\,  - p_2(x) \,   p_3(x)^{3} \, \,  
+ p_3(x)^{2} \, p_1(x) \, \,  -8\, p_0(x)  p_3(x) 
 \nonumber \\
\hspace{-0.95in}&& \quad   \qquad \qquad 
-4\, p_2(x)\,   p_1(x) \,  \,  \, 
+4\,  {\frac {d p_3(x)}{dx}} {\frac {d^{2} p_3(x)}{d{x}^{2}}}   \,  \, 
-4\, p_2(x)\,  {\frac {d^{2} p_3(x)}{d{x}^{2}}}   \,  \, 
+4\,{\frac {d^{2} p_1(x)}{d{x}^{2}}}
 \nonumber \\  
\hspace{-0.95in}&& \quad   \qquad \qquad
 +6\, ({\frac {d p_3(x)}{dx}})^{2} \,  p_3(x)  \,  \, 
+ p_3(x)^{3} \, {\frac {d p_3(x)}{dx}}   \,  \, 
-10\, p_3(x) \,  p_2(x) \, {\frac {d p_3(x)}{dx}} \, 
  \nonumber \\
\hspace{-0.95in}&& \quad   \qquad \qquad 
+8\, {\frac {d p_3(x)}{dx}} \,  p_1(x)  \,  \,  \, 
-8\, {\frac {d p_3(x)}{dx}} {\frac {d p_2(x)}{dx}}  \,  \,   \, 
+8\, p_2(x) {\frac {d p_2(x)}{dx}}
  \nonumber \\
\hspace{-0.95in}&& \quad  \qquad \qquad 
+6\, p_3(x) \,  {\frac {d p_1(x)}{dx}}  \,  \,   \, 
-8\,{\frac {d p_0(x)}{dx}}, 
\nonumber 
\end{eqnarray}
and:
\begin{eqnarray}
\label{CYext}
\hspace{-0.95in}&& \quad   \quad 
C_1^{(4)}\,  \,=\, \,\, \, \,  
4\,p_2(x)\, \,  p_3(x) \, \,  \,  -8\,p_1(x)\,  \, \,   -  p_3x)^{3} \,  \, 
+8\,{\frac {d p_2(x)}{dx}}\,  \, 
 -4\,{\frac {d^{2} p_3(x)}{d{x}^{2}}}
 \nonumber \\
\hspace{-0.95in}&& \quad   \qquad \qquad \, \,  \, 
-6\,p_3(x) \,  {\frac {d p_3(x)}{dx}}. 
\end{eqnarray}
Here also, $ \, C_1^{(4)} = \, 0$ is the so-called~\cite{2013-green-sg}
 ``Calabi-Yau condition''. When verified, $ \, Ext^2(L_4)$ is of
order five, instead of the order six.

All what have been said for $ \, L_3$ (on the 
rationality of the coefficients) 
holds. In particular, (\ref{thea0xE}) can be reduced
to the inhomogeneous linear differential equation
\begin{eqnarray}
\hspace{-0.95in}&& \quad  \quad  \quad \qquad   
2\, {\frac {d a_0(x)}{dx}} 
\,\,  +  p_3(x) \cdot \, a_0(x)\, \, \,  =\, \,\,  \, 
E_3 \cdot \, a_2(x)
\end{eqnarray}
with:
\begin{eqnarray}
\hspace{-0.95in}&& \quad   \qquad
 E_3  \, \, =\,\, \,  \, 
 D_x^3 \,\,\,  +2\, p_3(x) \cdot \, D_x^2 \, \, \,
+\, (p_3(x)^2\, +p_2(x)) \cdot \, D_x\, 
 \nonumber \\
\hspace{-0.95in}&& \quad \quad   \qquad \quad 
+p_2(x)\, p_3(x) \, \, \,  -p_1(x) \,\, \,  +2 {\frac {d p_2(x)}{dx}}\,\,  \, 
-2\, {\frac {d^2 p_3(x)}{dx^2}} \, \,\,  \, 
-p_3(x) \,  {\frac {d p_3(x)}{dx}}.
 \nonumber
\end{eqnarray}

\vskip 0.1cm

Similarly to $\, L_3$, one may consider an order-zero
 intertwiner $ \, a_0(x)$ for $ \, R_2$.
In this case, one obtains
\begin{eqnarray}
\hspace{-0.95in}&&  \quad \quad \quad \quad \qquad  
a_0(x) \, \, = \,\,\,  W_{L_4}(x)^{1/2} \,\, 
\qquad {\rm and} \,\, \qquad \quad  C_1^{(4)} \,=\,\,0.
\end{eqnarray}
$\, W_{L_4}(x)$ is the Wronskian of $\,L_4$ and $\,C_1^{(4)} = \,0$ is 
the Calabi-Yau condition~\cite{2013-green-sg}
given in (\ref{CYext}), which if fulfilled,
 $\,L_4$ has an order-five exterior square
instead of order six.

\subsection{Order-four differential operator and $\, SO(4, \,\mathbb{C})$}
\label{genericL4R31}

We consider the equivalence relation
\begin{eqnarray}
\hspace{-0.95in}&& \quad   \quad   \qquad  
 L_4 \cdot \,  R_3 \,\,\,  = \, \,\, \, adjoint(R_3) \cdot\,  adjoint(L_4), 
\end{eqnarray}
with $ \, R_3$ of order three
\begin{eqnarray}
\hspace{-0.95in}&& \quad  \quad   \qquad  
R_3 \, \,\,  \,  = \, \,\, \,  \, 
 a_3(x) \cdot \, D_x^3 \,\, \, \, + a_2(x)  \cdot \,D_x^2 \, \,\, \,
 + a_1(x) \cdot \, D_x \, \,\,  \, + a_0(x).
\end{eqnarray}

One obtains
\begin{eqnarray}
\label{a3symorderfour}
\hspace{-0.95in}&& \, \quad    
a_3(x)\,  \,=\,\, \,  sol(Sym^2(L_4)),
 \quad \quad   
a_2(x) \, \,=\, \, \, 
-p_3(x) \cdot \, a_3(x)\, \, 
-{1 \over 2}\,{\frac {d a_3(x)}{dx}}.
\end{eqnarray}

The coefficient $\, a_1(x)$ is a solution of the order-three 
inhomogeneous differential
equation \footnote[2]{Note that $\, a_1(x)$ is also given
 by $\,a_1(x) \, = \,\,  N_9 \cdot \, a_3(x)$, where
$\,N_9$ is an order-nine linear differential operator. Once $\, a_3(x)$
 is rational, $\, a_1(x)$ 
will be rational.}
\begin{eqnarray}
\hspace{-0.95in}&& \quad  \quad  \quad   \qquad  
N_3 \cdot\,  a_1(x) \, \, = \,\,\,  \,   N_5 \cdot\,  a_3(x), 
\end{eqnarray}
where the order-three operator $\, N_3$ is
\begin{eqnarray}
\hspace{-0.95in}&& \quad  \quad \, \,  
N_3 \,\, \,  =\,\,\,   \, 
5\,D_x^3 \, \, \,  +{15 \over 2}\, p_3(x) \cdot \,  D_x^2 \, \,  \, 
+ \left( {9 \over 2}\,{\frac {d p_3(x)}{dx}}\,   
+3\, p_3(x)^{2} \, +2\, p_2(x)  \right) \cdot \,   D_x 
\nonumber \\
\hspace{-0.95in}&& \quad   \quad  \qquad \qquad \quad \,  \, 
+ \left( 3\,{\frac {d p_2(x)}{dx}} \,  -2\, p_1(x) \, 
+2\, p_2(x)\,   p_3(x)  \right),  
\end{eqnarray}
and the order-five operator $\, N_5$ reads
\begin{eqnarray}
\hspace{-0.95in}&& \quad   \quad \quad 
N_5\, \, \, =\,\,\,\, 
 D_x^5\,\, \,  +{\frac{15}{4}}\, p_3(x) \cdot \, D_x^4 \,\,  \, 
+ \left( {\frac {17}{4}}\, p_3(x)^{2}
-{7 \over 4}\,{\frac {d p_3(x)}{dx}} \, 
 + {9 \over 2}\, p_2(x)  \right) \cdot \, D_x^3 
\nonumber \\
\hspace{-0.95in}&& \quad  \quad  \qquad \qquad \quad
+ Q_2 \cdot \, D_x^2 \,\,\, \,  + Q_1 \cdot \, D_x \,\,  \,\, + Q_0, 
\end{eqnarray}
with:
\begin{eqnarray}
\hspace{-0.95in}&& \quad   \quad
Q_2 \,\, = \, \, \, 
{3 \over 2}\, p_3(x)^{3} \, \, \, 
+{17 \over 2}\, p_2(x) \,  p_3(x)  \, \, \, 
-{5 \over 2}\, p_1(x)  \, \, \, 
-{\frac {45}{2}}\,{\frac {d^{2} p_3(x)}{d{x}^{2}}} 
  \nonumber \\
\hspace{-0.95in}&& \quad   \quad \qquad \quad 
-{\frac {21}{4}}\, p_3(x) {\frac {d p_3(x)}{dx}} \,  \, \, 
+15\,{\frac {d p_2(x)}{dx}},  
\end{eqnarray}
\begin{eqnarray}
\hspace{-0.95in}&& \quad   \quad
Q_1 \,\, = \, \, 
2\, p_2(x)^{2} \, -2\, p_1(x) \,  p_3(x)  \,\,  \, 
+4\, p_2(x) \, p_3(x)^{2} \, -4\, p_0(x)\,\, \, 
-{\frac {27}{4}} \, ({\frac {d p_3(x)}{dx}})^{2}
 \nonumber \\
\hspace{-0.95in}&& \quad   \quad \qquad \quad 
-{\frac {105}{4}}\, p_3(x)\, {\frac {d^{2} p_3(x)}{d{x}^{2}}}  \,\, 
-{9 \over 2}\, p_3(x)^{2}\, {\frac {d p_3(x)}{dx}}   \, \, 
+{3 \over 2}\, p_2(x)\, {\frac {d p_3(x)}{dx}}  
 \nonumber \\
\hspace{-0.95in}&& \quad   \quad \qquad \quad 
+{\frac {33}{2}}\, p_3(x)\, {\frac {d p_2(x)}{dx}}   \, \, 
+15\,{\frac {d^{2} p_2(x)}{d{x}^{2}}}  \, \, 
-{\frac {55}{2}}\,{\frac {d^{3} p_3(x)}{d{x}^{3}}}   \,  \,  
-{3 \over 2}\,{\frac {d p_1(x)}{dx}},  
\end{eqnarray}
\begin{eqnarray}
\hspace{-0.95in}&& \quad   \quad
Q_0   \, \,=\,  \,   \,  
2\, p_3(x)   \,  (p_2(x))^{2} \, 
-2\, p_0(x)   \,  p_3(x) \, \, 
 -2\, p_1(x)  \,   p_2(x)  \, \, 
-10\,{\frac {d^{4} p_3(x)}{d{x}^{4}}}
 \nonumber \\
\hspace{-0.95in}&& \quad   \quad \qquad \quad 
-3\,{\frac {d p_0(x)}{dx}} \,\,  
 -15\, p_3(x)   \, {\frac {d^{3} p_3(x)}{d{x}^{3}}}  \,   \, 
-{3 \over 2}\,  {\frac {d p_2(x)}{dx}}  \, {\frac {d p_3(x)}{dx}} \,   \, 
+5\, p_2(x) {\frac {d}{dx}} p_2(x) 
\nonumber \\
\hspace{-0.95in}&& \quad   \quad \qquad \quad 
-9\, {\frac {d p_3(x)}{dx}} {\frac {d^{2} p_3(x)}{d{x}^{2}}}   \, \, 
+4\, p_1(x)\,  {\frac {d p_3(x)}{dx}}     \, \, 
-4\, p_2(x) \, {\frac {d^{2} p_3(x)}{d{x}^{2}}}  \,\,  
+5\,{\frac {d^{3} p_2(x)}{d{x}^{3}}}
  \nonumber \\
\hspace{-0.95in}&& \quad   \quad \qquad \quad 
+{15 \over 2}\, p_3(x) \,  {\frac {d^{2} p_2(x)}{d{x}^{2}}}   \,\,  
-6\, p_3(x)^{2} \, {\frac {d^{2} p_3(x)}{d{x}^{2}}}  \,  \, 
-4\, p_2(x)\,  p_3(x)\, {\frac {d p_3(x)}{dx}} 
 \nonumber \\
\hspace{-0.95in}&& \quad   \quad \qquad \quad 
+3\, {\frac {d p_2(x)}{dx}} \,\,  \, p_3(x)^{2}.
\end{eqnarray}

The coefficient $ \, a_0(x)$ is given by
\begin{eqnarray}
\hspace{-0.95in}&& \quad   \quad \qquad  
a_0(x)\,\,  \,=\,\,\,\, 
 -p_3(x) \cdot \, a_1(x)\, \,\, 
 -{3 \over 2}\,{\frac {d a_1(x)}{dx}}  \,\,\, 
+{1 \over 4}\, E_3 \cdot \, a_3(x), 
\end{eqnarray}
where the order-three operator $\,E_3$ reads:
\begin{eqnarray}
\hspace{-0.95in}&& \,\,\, 
E_3\, \,\, =\, \,\,\, 
D_x^{3} \, \, +3\, p_3(x)  \cdot  \,  D_x^{2} \, \,\, 
+ \left( 4\, p_2(x) \,
-7\,{\frac {d p_3(x)}{dx}} \,  
+2\,   p_3(x)^{2} \right) \cdot  \,  D_x 
\nonumber \\
\hspace{-0.95in}&&   \quad  \, \, 
+4\, p_2(x) \, p_3(x)\, \, -4\, p_1(x)  \,\, 
-16\,{\frac {d^{2} p_3(x)}{d{x}^{2}}}\,\, 
\,  +10\,{\frac {d p_2(x)}{dx}}  \,\, 
-8\, p_3(x)\, {\frac {d p_3(x)}{dx}}. 
\end{eqnarray}

For the equivalence, between the differential operator 
$ \, L_4$ and its adjoint with
an order-three intertwiner {\em over the rationals}, to exist,
 it is the {\em symmetric square} 
of $ \, L_4$ that should annihilate a rational solution.

\vskip 0.2cm

We consider, now, the case of an order-one intertwiner 
$ \, R_1 =\, \,  a_1(x) \cdot \,D_x  \, + a_0(x)$ in 
\begin{eqnarray}
\hspace{-0.95in}&&   \quad  \quad \quad \qquad 
 L_4 \cdot \,R_1 \,\, \, =\,\,\,  \, adjoint(R_1) \cdot\, adjoint(L_4).
\end{eqnarray}
One obtains for $\, a_0(x)$
\begin{eqnarray}
\hspace{-0.95in}&& \quad \,\qquad \qquad 
a_0(x) \,\, \, = \,\,\, \,
 -p_3(x) \cdot \, a_1(x)\, \,  \,
 -{3 \over 2}\,{\frac {d a_1(x)}{dx}}, 
\end{eqnarray}
and
\begin{eqnarray}
\hspace{-0.95in}&& \quad \,\qquad \qquad 
a_1(x) \,\,\, =\,\,\,\, sol(E_2), 
\end{eqnarray}
where the order-two linear differential operator $\,E_2$ reads:
\begin{eqnarray}
\label{E2fora1x}
\hspace{-0.95in}&&  \quad 
E_2 \, \,\, = \, \,\,\,
C_1^{(4)} \cdot\, D_x^2\,\,
 + {1 \over 105}\, Q_1 \cdot \, D_x \,\, + {2 \over 105}\, Q_0, 
\end{eqnarray}
\begin{eqnarray}
 \hspace{-0.95in}&& 
Q_1 \,\,  =\, \, \, 
212\,\, p_2(x)  \,  p_3(x)^{2}\,  \, 
-66\, p_3(x)^{4}\,  -440\, p_1(x) \,  p_3(x) \,  \, 
+144\, p_2(x)^{2} \, -1600\, p_0(x), 
\nonumber \\
 \hspace{-0.95in}&& \quad \quad \quad \, 
+54\, ({\frac {d p_3(x)}{dx}})^{2}\,  \, 
-292\, p_2(x) {\frac {d p_3(x)}{dx}}  \,  \, 
-60\,{\frac {d^{3} p_3(x)}{d{x}^{3}}} \,   \, 
-40\,{\frac {d^{2} p_2(x)}{d{x}^{2}}} \, 
  \nonumber \\
 \hspace{-0.95in}&& \quad \quad \quad 
+520\,{\frac {d p_1(x)}{dx}}  \, \,   
+580\,  {\frac {d p_2(x)}{dx}} \,   p_3(x) \,  \, 
-423\, p_3(x)^{2} \, {\frac {d p_3(x)}{dx}} \,   \, 
-390\, p_3(x) \,  {\frac {d^{2} p_3(x)}{d{x}^{2}}}, 
\nonumber
\end{eqnarray}
\begin{eqnarray}
 \hspace{-0.95in}&& 
Q_0 \,\,  =\, \, \, 
-22\, p_2(x)  \,   p_3(x)^{3}\,  \, 
+22\, p_1(x) \,    p_3(x)^{2}\,  \, 
-72\, p_1(x)\,   p_2(x) \,  \, 
+72\, p_2(x)^{2} \,  p_3(x)
   \nonumber \\
 \hspace{-0.95in}&& \quad \quad \quad 
-400\, p_0(x)\,   p_3(x)  \,  \, 
-13\,  {\frac {d p_2(x)}{dx}}  \,  \, 
   p_3(x)^{2} \,  \, 
+18\, {\frac {d p_2(x)}{dx}} \,  \, 
{\frac {d p_3(x)}{dx}}  \, 
-20\,{\frac {d^{3} p_2(x)}{d{x}^{3}}}  
 \nonumber \\
\hspace{-0.95in}&& \quad \quad \quad 
-50\,  {\frac {d^{2} p_2(x)}{d{x}^{2}}} \,   p_3(x) \, \,
+180\,  {\frac {d p_1(x)}{dx}} \,    p_3(x) \, \,
-52\, p_1(x) \,  {\frac {d p_3(x)}{dx}}  
  \nonumber \\
 \hspace{-0.95in}&& \quad \quad \quad 
-128\, p_3(x)\,   p_2(x) \,  {\frac {d p_3(x)}{dx}}  \, \,
  -200\,{\frac {d p_0(x) }{dx}} \, \,
+80\,{\frac {d^{2} p_1(x)}{d{x}^{2}}} \, \,
-80\, p_2(x) \, {\frac {d^{2} p_3(x)}{d{x}^{2}}}
   \nonumber \\
 \hspace{-0.95in}&& \quad \quad \quad 
+108\,  {\frac {d p_2(x)}{dx}} 
 \,   p_2 \left( x \right). 
 \nonumber
\end{eqnarray}

Here, our analysis is a little bit incomplete. For all the examples so far found, this
situation corresponds to a symmetric square of order nine instead of the order ten.
We do not know how to prove that (\ref{E2fora1x}) should have a rational solution
in this case. One may just imagine, that this is provable 
if we use the symmetric Calabi-Yau
condition (the corresponding condition for order-four 
operators that makes the symmetric square
of order nine instead of ten) to factorize, or reduce to order 
one, the operator (\ref{E2fora1x}). 
Unfortunately, this condition for order-four operator is 
very large, and nonlinear, in the
coefficients of $ \, L_4$ and their derivatives: it is sum 
of 3548 monomials and has degree 12 in 
the coefficients (and their derivatives) of $ \, L_4$.
 
\section{Quadratic (or alternating) invariant forms}
\label{invariantform}

The differential Galois group of the
order-three linear differential operator $ \,F_3$, 
occurring in $ \,\tilde{\chi}^{(5)}$, is in
the orthogonal group $ \,SO(3, \,\mathbb{C})$.
We denote by $ \,X_0$ any of its formal solutions 
at $\,x= \,0$ and by $ \,X_1$ (and $ \,X_2$) 
respectively the first (and second) derivative. The first 
integral of $ \,F_3$ reads
\begin{eqnarray}
\hspace{-0.95in}&& \quad  \quad  \quad  \quad 
Q( X_0, \, X_1,\, X_2)\, \, = \, \, \,const.
\end{eqnarray}
with
\begin{eqnarray}
\hspace{-0.95in}&& \, \,    Q(X_0,\, X_1,\, X_2) \, \, \,  = \, \, \,\, \, 
 {\frac{x \cdot \, (1+4x)^2(1-4x)^6
  \cdot \, P_{101}(x)}{A_7(F_2)^4  \cdot \, A_{37}(F_3)^2}}\cdot \, X_0^2 
 \nonumber \\
\hspace{-0.95in}&& \quad 
+{\frac{x^3 \cdot \,(1+4x)^4(1-4x)^8(1+2x)^2 \cdot  
\, P_{81}(x)}{A_7(F_2)^2  \cdot \, A_{37}(F_3)^2}}\cdot \, X_1^2 
 \nonumber \\
\hspace{-0.95in}&& \quad  
+{\frac{x^5  \cdot \, (1+4x)^6 \, (1-4x)^{10} \, (1-x)^2 \,
 (1+2x)^4 \, (1-2x)^2 \, (1+3x+4x^2)^2 \cdot \, P_{53}(x)}{A_{37}(F_3)^2}}
 \cdot \, X_2^2
 \nonumber \\
 \hspace{-0.95in}&& \quad 
-{\frac{x^4 \cdot \,(1+4x)^5 \, (1-4x)^9 \,(1-x) \,(1+2x)^3 \,
(1-2x) \,(1+3x+4x^2) \cdot \, P_{67}(x)}{A_7(F_2)  \cdot \, A_{37}(F_3)^2}}
\cdot \, X_1 X_2
 \nonumber \\
\hspace{-0.95in}&& \quad  
+{\frac{x^3  \cdot \,(1+4x)^4 \,(1-4x)^8 \,(1+2x)^2 \, (1-x) \,
 (1-2x) \,(1+3x+4x^2) \cdot \, P_{77}(x)}{A_7(F_2)^2  \cdot \, A_{37}(F_3)^2}} 
\cdot \, X_0\, X_2 
\nonumber \\
 \hspace{-0.95in}&& \quad 
-{\frac{x^2 \cdot \, (1+4x)^3 \,(1-4x)^7 \,(1+2x)
 \cdot \, P_{91}(x)}{A_7(F_2)^3 \cdot \, A_{37}(F_3)^2}} \cdot \,  X_0 \, X_1. 
\end{eqnarray}
The $P_j(x)$'s are polynomials of degree $j$.
The numerical value of $ \, const.$ depends on the solution $\, X_0$ considered.

\vskip 0.2cm

The differential Galois group of the order-four linear differential operator $\,L_4$,
 occurring in $\,\tilde{\chi}^{(6)}$, is in
the symplectic group $\, Sp(4, \,\mathbb{C})$.
We call $\,X_0$ any of its formal solutions at $\,x=\,0$ and $\,X_j$, the $j^{th}$
 derivative up to $j=\,3$. 
$\,Y_0$ is another solution with its derivatives $\,Y_j$, $j=\,1,\, \cdots,\, 3$. We define
\begin{eqnarray}
\hspace{-0.95in}&& \quad  \quad 
w_{i, j} \,\,  =\,\, \,  X_i \cdot Y_j - X_j \cdot Y_i, \qquad i\,=\,\,0, 
\cdots, 3, \,\, \quad j=\,0, \,\cdots, \,3, \quad \, j > i.
\end{eqnarray}
The first integral of the order-four differential operator $\, L_4$ reads
\begin{eqnarray}
\hspace{-0.95in}&& \quad \quad \qquad 
Q(X_0,\, X_1,\, X_2,\, X_3,\, Y_0, \,Y_1, \,Y_2,\, Y_3) \,\,=\,\, \,const.
\end{eqnarray}
where
\begin{eqnarray}
\hspace{-0.95in}&& \quad \quad    Q\,\,  \, =\, \,\,\,\, 
 {\frac{x^7 \,(1-16 x)^{10} \cdot \, P_{36}}{A_{26}(L_4) A_4(\tilde{L}_3)^3}} \cdot \, w_{0,1}
\,\,\,
 +{\frac{x^8 \cdot \, (1-16 x)^{11}\, (1-8 x) \cdot \,
 P_{34}}{A_{26}(L_4) A_4(\tilde{L}_3)^3}} \cdot \, w_{0,2}
\nonumber \\
\hspace{-0.95in}&& \quad  \quad  \qquad \qquad  \, 
 + {\frac{x^9 \cdot \,(1-16 x)^{12} \, (1-4 x) (1-8 x) \cdot \, 
P_{28}}{A_{26}(L_4)\cdot  \, A_4(\tilde{L}_3)^2}} \cdot  \, w_{0,3}
\nonumber \\
\hspace{-0.95in}&& \quad  \quad  \qquad \qquad 
\, + {\frac{x^9 \cdot \,  (1-16 x)^{12} \,(1-8 x) 
\cdot \, P_{29}}{A_{26}(L_4) \cdot  \,A_4(\tilde{L}_3)^2}} \cdot \, w_{1,2}
 \nonumber \\
\hspace{-0.95in}&& \quad  \quad \qquad \qquad 
 \, + {\frac{x^{10} \cdot \,(1-16 x)^{13}\, (1-4 x)\, (1-8 x)
 \cdot \, P_{23}}{A_{26}(L_4) \cdot  \,A_4(\tilde{L}_3)}} \cdot  \, w_{1,3}
\nonumber \\
\hspace{-0.95in}&& \quad   \quad \qquad \qquad 
\, + {\frac{x^{11}\, (1-16 x)^{14} \,(1-4 x)^2\, (1-8 x)
 \cdot   \,P_{17}}{A_{26}(L_4) }} \cdot \, w_{2,3}. 
\nonumber
\end{eqnarray}
$P_j(x)$ are polynomials of degree $j$.
The numerical value of $\, const.$ depends on the 
solutions $\, X_0$ and $ \, Y_0$ considered.

\section{Minimal order versus non minimal order: $ \, F_3 \cdot F_2$}
\label{appendixF3F2}

For a given series $\, S(x)$ solution of a Fuchsian operator of order $q$, one can consider 
the family of {\em Fuchsian} linear differential operators of order $\, Q > q$ 
annihilating this series, the degree $\, D$ of the polynomial coefficients 
being taken as small as possible~\cite{2008-experimental-mathematics-chi}. 
The Fuchsian operator of minimal order
annihilating this series is unique, and rightdivides all the operators 
of the previous family. Let us denote by $\, N$ the minimum number of coefficients needed to find the linear ODE 
in this family within the constraint that the order $Q$ and degree $D$ 
are given. We found (as an experimental result) that $\, N$, 
the order $Q$ and the degree $D$ of the operators in this family are related by 
a {\em linear relation}, we called an "ODE formula''
 (see section 3.1 of~\cite{2008-experimental-mathematics-chi}).

Assume we have a series $\, S(x)$, known modulo a prime,
for which we have produced an ODE whose
 "ODE formula~\cite{2008-experimental-mathematics-chi}" reads
\begin{eqnarray}
\label{ODEformula}
\hspace{-0.95in}&& \quad   \quad \qquad
N \,  \, =\, \, \,
14 \, Q \, \, + 5 \, D \, \, -14
\, \,  \, =\, \, \,  \, (Q+1) \cdot \, (D+1)\, \, \, -f, 
\end{eqnarray}
where $\, Q$ is the order and $\, D$ is the degree of
 $\, P_j(x)$, $j=\, 0, \cdots, Q$ of the ODE written in the form
$\, P_Q(x)\cdot\,  x^Q \cdot\,  D_x^Q \, + \cdots$ + $\, P_0(x)$,
and $\, f$ is the number of the independent non-minimal order ODE
with $Q$ and $D$ such that $(Q+1)(D+1) <$ the available number of the series
terms of $S(x)$ (see~\cite{2008-experimental-mathematics-chi} 
and Appendix B of~\cite{2009-chi5} for the details). From the 
"ODE formula~\cite{2008-experimental-mathematics-chi}" we see that
the minimal order ODE annihilating $\, S(x)$ is of order 5, 
and we call it $\, {\cal F}_5$. 
Among the many non minimal order ODEs there is one which 
needs the lesser terms in $\, S(x)$ to be produced. This particular 
{\em non minimal order} linear ODE is the "optimum ODE", and has, for this 
example, the order eight, i.e. $(Q_0=\, 8, \, D_0=\, 23,\,  f_0=\, 3)$. 
In the calculations we may use this order-eight ODE
 and continue to call it $\, {\cal F}_5$,
but in this section we call it $\, {\cal F}_5^{nm}$. Obviously,
 $\, {\cal F}_5$ is a right factor of $\, {\cal F}_5^{nm}$.

The local exponents at $\, x=\, 0$ of this linear 
ODE are $\, 0,\,  0,\,  1, \, 2,\,  3$. These
 exponents are obtained
whatever the order of the ODE is, in minimal order (i.e. ${\cal F}_5$), or
 in non minimal order (i.e. ${\cal F}_5^{nm}$). 
The three exponents corresponding to the three extra solutions
 of the order-eight linear ODE
${\cal F}_5^{nm}$ appear as {\em non rational numbers}. Recall
 that we are dealing with globally nilpotent
differential equations~\cite{2009-global-nilpotence}. 
They are therefore necessarily Fuchsian and have 
{\em rational local exponents at all the singular points}.

The general (analytic at $x= \, 0$) solution $\, \tilde{S}(x)$ 
of $\, {\cal F}_5$ (or $\, {\cal F}_5^{nm}$)
depends on two free coefficients, say $\, \alpha$ and $\, \beta$. 
The series $\, S(x)$ is a particular combination of $\tilde{S}(x)$. 
With $\, Q=\, Q_0=\, 8$, $\, D=\, D_0=\, 23$, if there are
 some values of $(\alpha= \, \alpha_0, \, \beta= \, \beta_0)$ 
in the range $\, [1,\,  p_r]$, for which $\, f$ is 
greater than $\, f_0=\, 3$, the linear differential 
operator $\, {\cal F}_5$ has a right factor. The solution $\, \tilde{S}(x)$ with 
$\, (\alpha=\, \alpha_0,\,  \beta=\, \beta_0)$ 
gives an ODE, whose "ODE formula" reads
\begin{eqnarray}
\hspace{-0.95in}&& \quad   \quad \qquad
7 \, Q\,  \, + 2 \, D \, \, +2 
\,\, \,  =\, \,  \, \, (Q+1)\cdot \,(D+1)\,\,  \, -f, 
\end{eqnarray}
telling that, indeed, there is an order-two right factor 
occurring in the linear differential 
operator $\, {\cal F}_5$. We call this factor $\, F_2$, when 
obtained as minimal order ODE,
and $\, F_2^{nm}$ if it is obtained in non minimal order.

\vskip 0.1cm

Here begin the details of our Remark 3. To obtain the factor (call it $\, F_3$) at 
the left of $\, F_2$ in $\, {\cal F}_5$, we may just use the "rightdivision" command of 
DEtools in Maple, or act by $ \, F_2$ on the series $\, S(x)$ to obtain a series and look 
for the linear ODE annihilating it.
But assume that this is cumbersome, or not doable. Either the "rightdivision" command
is not feasible, or the series $\, F_2(S(x))$ has no more enough coefficients terms
to encode the remaining left factor. This what happens in the case of
 $\, L_{21} \cdot\,  \tilde{L}_2$
when $\, \tilde{L}_2$ is of minimal order.

Let us give some details on the factorization
 $\, {\cal F}_5 =\,  F_3 \cdot \, F_2$.
Assume we have obtained $\, F_3$ from the series $\, F_2^{nm}\left( S(x) \right)$.
The rational solution of the symmetric square of $\, F_3$
 (in minimal order or in non minimal order)
will appear as
\begin{eqnarray}
{\frac{P_{32}(x)}{(1-4x)^7 \, (1+4x)^7\,  (1-2x)\,  (1+2x) }}.
\end{eqnarray}
If we use the minimal order $\, F_2$ to obtain $\, F_3$.
The rational solution of the symmetric square 
of $\, F_3$ (in minimal order or in non minimal order) will appear as
\begin{eqnarray}
\label{RatSolsym2F3appendix}
{\frac {P_{34}(x)}{ (1-4\,x)^{5}\,  \, (1+4\,x)^{5} }}.
\end{eqnarray}
To obtain the rational solution given in (\ref{RatSolsym2F3}), one has to divide by
the coefficient of the higher derivative of $ \, F_2$. This 
is because, we have used $\, F_2$,
in non monic form, to mimic the situation of the large orders linear 
differential operators
for which the non monic form is more tractable in the computations.

As far as the occurrence of a rational solution to the 
symmetric square of a left factor
is concerned, it is irrelevant whether the right factor
 is of minimal order or in non minimal order.

\end{document}